%% file: main.tex
\documentclass[11pt]{article}
\usepackage[margin=1in]{geometry}
\usepackage{amsmath,amssymb}
\usepackage{booktabs}
\usepackage{graphicx}
\usepackage{xcolor}
\usepackage[hidelinks]{hyperref}
\usepackage{natbib}
\usepackage{microtype}
\usepackage{url}

\input{tables/numbers.tex}

\title{Reachability, Not Observation:\\
Containing Systems Whose Wiring Changes}
\author{Yoshiaki Takashita\thanks{This work was carried out independently and is not affiliated with any laboratory.}\\
  School of Law, Waseda University\\
  takashita@moegi.waseda.jp}
\date{September 2026}

\begin{document}
\maketitle

\begin{abstract}
Containment decisions---where to place a firewall, which links to monitor, what a program may reach---are made from an observed structure, and observation is a snapshot.
We ask what a snapshot misses when the wiring changes over time, and answer with three systems built by the author.
First, a hypercube whose active dimension rotates from step to step.
With instantaneous degree one it reaches every node in \nRotSteps{} steps, where the full hypercube of degree \nDBig{} takes \nCubeSteps{}; a balanced split shows zero crossing edges at \nRotZero\% of instants yet requires \nRotTemporal{} edges to be blocked permanently.
Add one always-on ring and a defender counting crossing edges sees \nRlInst{} where \nRlTemporal{} must be blocked, a factor of \nRlRatio{}.
The same budget of monitoring taps sees \nRingTaps\% of traffic on a ring and \nRotTaps\% on rotating wiring.
A defender who models time, however, holds on average \nRotAware{} blocks, \nAwareRatio{} times fewer than a static hypercube demands, provided its clock is exact: one step of lag drops containment to \nLagOneHeld\%.
Rotation does not raise the cost of containment, it punishes containment computed from a snapshot.
Second, the Internet: on \nAsDays{} daily snapshots of the autonomous-system graph from \nAsFirst{} to \nAsLast{}, the same gap over two years is $\times$\nAsRewireLo--\nAsRewireHi{} once growth is removed, and on \nCaMonths{} monthly snapshots from \nCaFirst{} to \nCaLast{} it is $\times$\nCaRewireLo--\nCaRewireHi{}.
The blind spot is therefore introduced by design, not inherited from the world; and it has been designed: the round-robin matching schedules of optical datacentre fabrics have a gap equal to their period, $n-1$, and are contained by no observation window shorter than the whole cycle.
Third, a declared capability map with a static reachability check over the call graph of a trading application: of \nGHoles{} holes planted deliberately, the reachability check catches \nGHoles{} and the string deny-list that was in place before catches \nGCaughtOld{}.
We read the three as one statement: contain by the paths that exist, not by the behaviour that was seen.
A short calculation (\S\ref{sec:closed}) shows that every number in the first part follows from one parameter, the period of the schedule, and reproduces the measurements exactly.
The same calculation reads three boundaries that are not usually called schedules, and one that is: frequency hopping, whose standard results (processing gain, wideband jamming, the follower jammer's reaction time) are the four closed forms with channels in place of edges.
An air gap is a schedule whose always-on crossing set is empty: media inspection is its time-aware cut, and the known breach of one was a crossing at the one phase a snapshot does not see.
The tool surface of a coding agent, inventoried from the inside, is \nAgVisiblePct\% visible at the start of a session in tools and fully visible in effects, because a shell is; removing every named tool for an effect leaves \nAgDenyStill{} of \nAgEffects{} effects reachable through it.
Finally we turn the cuts on the agent itself.
Its instantaneous state cut while idle is zero; its temporal cut is \nAmChannels{} channels through which it survives a context reset, none of which is the network, so severing the network removes \nAmNetLost{} of them.
One of those channels spawns agents with the same tools, and treated as a branching process it has a sharp threshold at approval rate $1/b$: below it denial is unnecessary, above it denial is insufficient, and only a finite activation budget contains it.
\end{abstract}

\section{Introduction}
\label{sec:intro}

A defender looks at a system as it is and decides where to cut.
The firewall goes on the links that cross between two zones; the monitoring taps go where the traffic is; the program is allowed to touch what it was seen to touch.
Every one of these decisions is computed from an observed structure, and an observation is a snapshot.
The question of this paper is what the snapshot leaves out when the structure it observed is not the structure that will exist a moment later.

We did not come to this question from security.
A companion paper \citep{takashita2026rotating} showed that a sparse wiring pattern on a hypercube, in which each layer connects a position only to its neighbour along one dimension and the dimension \emph{rotates} from layer to layer, reaches every position in $\log_2 n$ layers with $2n$ links per layer, and that this is enough to replace most of a language model's attention layers at a fraction of the cost.
The property that makes the pattern useful is that every node reaches every other with almost no links present at any one time.
Read from the other side, that is a description of a system that cannot be contained by looking at it.

This paper measures that other side.
It has three parts, and the point is the way they fit.

\paragraph{Claims.}
(N) On rotating hypercube wiring, the number of edges a defender sees crossing a balanced split at a random instant is zero \nRotZero\% of the time, while \nRotTemporal{} edges must be blocked to keep the halves apart over one period; with one always-on ring added, the visible count is \nRlInst{} and the true count \nRlTemporal{} (\S\ref{sec:snapshot}).
A defender who accounts for time needs on average \nRotAware{} blocks, \nAwareRatio{} times fewer than a static hypercube of the same reach, so the gap is a property of snapshot reasoning, not of the wiring; but that defender must know the schedule exactly, since a clock lag of one step drops its containment from \nLagZeroHeld\% to \nLagOneHeld\%, and a defender who merely watches for $w$ steps is protected with probability $w/d$ at a cost proportional to $w$ (\S\ref{sec:snapshot}).
The same holds for monitoring: a fixed budget of taps sees \nRingTaps\% of traffic on a ring and \nRotTaps\% on rotating wiring, independent of how packets choose their route (\S\ref{sec:taps}).
(R) On the real Internet the gap between one snapshot and two years of snapshots is at most $\times$\nAsRewireHi{} in 1997--2000 and $\times$\nCaRewireHi{} in 2024--2026, once growth is separated from rewiring (\S\ref{sec:wild}).
The gap in the rotating design is three orders of magnitude larger, and present from the first instant; in the round-robin schedules that optical datacentre fabrics actually run it equals the period $n-1$, and a defender that watches all but one phase contains nothing (\S\ref{sec:rotor}).
(G) A capability map declared in a small notation, checked by static reachability over the call graph of a real application, catches every one of \nGHoles{} planted holes; the string-matching test that was in place before caught \nGCaughtOld{} (\S\ref{sec:tool}).
The tool is the constructive half of the argument: it computes containment over the paths that exist rather than the behaviour that was observed.
(S) The calculation of \S\ref{sec:closed} is frequency hopping with time in place of frequency, and its four items are that field's four standard results (\S\ref{sec:hopping}); it also applies to two boundaries that are not built as schedules: an air gap, whose transfer events are its one-phase set, and the tool surface of a coding agent, where an inventory taken from the inside finds \nAgVisiblePct\% of tools visible at the start of a session but every effect reachable, and a deny-list that removes every named tool for an effect leaves \nAgDenyStill{} of \nAgEffects{} effects reachable through a shell (\S\ref{sec:single}).

(T) Turned on the agent itself, the two cuts say that its boundary is temporal: idle, it has no state to cut, and the \nAmChannels{} channels by which it reaches its next activation do not include the network (\S\ref{sec:amnesia}); the one channel that reproduces it is a branching process whose containment is decided by the approval \emph{rate} at threshold $1/b$, not by any approval, with a finite activation budget as the only lever above threshold (\S\ref{sec:spawn}).

Along the way we report where our own measurements misled us and how we found out (\S\ref{sec:negative}), because in the companion paper that turned out to be the part readers trusted most.

\section{Related work}
\label{sec:related}

\paragraph{Temporal networks and temporal cuts.}
Graphs whose edges are present only at certain times are studied under the name temporal networks \citep{holme2012temporal}; the notion of a time-respecting path, and the observation that reachability in such graphs is not the reachability of any static projection, goes back at least to \citet{kempe2002connectivity}, and the failure of Menger's theorem in scheduled networks to \citet{berman1996vulnerability}.
Separating two vertices of a temporal graph by deleting few vertices or time-labelled edges is NP-hard in general \citep{zschoche2020separators}, and remains so when the adversary may also perturb the remaining labels \citep{enright2026robust}.
Our two cuts (\S\ref{sec:cuts}) are the smallest instance of that theory applied to a defender: the partition is given, the edges are deleted permanently, and nothing has to be optimised, which is why our closed forms are lines of counting rather than algorithms.
The instantaneous cut is a static projection, the temporal cut is not.
We add nothing to the theory; we measure the size of the discrepancy for one engineered family and for the Internet, and then find the same discrepancy in three places that are not described as temporal graphs (\S\ref{sec:rotor}, \S\ref{sec:single}).
On the immunisation side, \citet{lee2012immunize} found that vaccinating a person's most recent contact, a window defender in our terms, is efficient on empirical contact networks; the reason is that human contacts persist, which is to say $|A|$ is large there.
Our rotation is the case $A = \emptyset$, where the same strategy is worth $w/T$ and no more.

\paragraph{Small worlds and wormholes.}
Adding a few long-range shortcuts to a lattice collapses its diameter \citep{watts1998collective}.
In wireless security the same idea appears as an attack: two colluding nodes tunnel packets between distant points, and the network's own routing believes they are adjacent \citep{hu2003packet}.
It is worth noting that in that literature the tunnel is the adversary's tool.
Our rotating wiring is a schedule of shortcuts rather than a fixed set of them, and the defender's difficulty is not that the shortcuts exist but that any one snapshot shows almost none of them.

\paragraph{Moving-target defence.}
A line of work proposes to change a system's configuration continually so that an attacker's reconnaissance goes stale \citep{jajodia2011moving}; \citet{zhuang2014theory} pose its timing problem (when to move) and \citet{hobson2014challenges} list, among the challenges of moving effectively, that movement also costs the defender.
Our measurements bear on this directly: the same change that invalidates the attacker's snapshot invalidates the defender's, and \S\ref{sec:snapshot} quantifies what the defender loses if it keeps reasoning from snapshots, and what an exact clock is worth.
We do not argue against moving targets; we argue that whoever moves them must also compute containment over time.

\paragraph{Reconfigurable datacentre networks.}
Optical datacentre fabrics that cycle through a fixed sequence of matchings on a schedule, so that every pair of racks is connected directly once per cycle, have been built \citep{mellette2017rotornet,ballani2020sirius}; Opera adds an always-on expander so that the instantaneous graph is connected at every moment while the union over a cycle is complete \citep{mellette2020opera}.
These are engineered rotation at production scale, and \S\ref{sec:rotor} measures the schedule they share.

\paragraph{Spread spectrum.}
A frequency-hopping radio is live on one channel per dwell and on every channel over a hop sequence; a receiver without the sequence sees a fraction of the signal equal to the inverse of the number of channels, and a jammer that follows the hops must react within the dwell \citep{pickholtz1982spread,torrieri1989repeater}.
\S\ref{sec:hopping} shows that this is \S\ref{sec:closed} with time in place of frequency, and reads our results, including the ones we found by simulation, as the standard results of that field.

\paragraph{Capability systems and information flow.}
Restricting what a component can reach by what it has been handed, rather than by what it is observed to do, is the capability discipline \citep{dennis1966programming,miller2006robust}, and the principle of least privilege is older still \citep{saltzer1975protection}.
Static analysis of what may flow where is the information-flow tradition \citep{denning1976lattice}.
The tool in \S\ref{sec:tool} is a small, practical instance: a declared map of who may reach what, checked against the call graph.
Its contribution is not the idea but the measurement of how much a string deny-list, the form such checks usually take in practice, actually misses.

\paragraph{Air gaps, covert channels and confinement.}
Physical isolation is the containment of last resort, proposed for artificial intelligence by \citet{yampolskiy2012leakproofing} and breached in practice by removable media \citep{falliere2011stuxnet,langner2011stuxnet} and by a long catalogue of covert emission channels \citep{guri2018bridgeware}.
That a confined program can leak through channels the confiner did not enumerate is the confinement problem as \citet{lampson1973confinement} stated it.
\S\ref{sec:airgap} does not extend that literature; it reads it through \S\ref{sec:closed}, in which an air gap is a schedule with an empty always-on set, and finds that its operational conclusions are the time-aware cut and the completeness rule of \S\ref{sec:tool}.

\paragraph{Agent tool surfaces.}
Language-model agents act through named tools with declared effects, increasingly discovered at run time over a protocol \citep{mcp2024}.
Restricting what such an agent may call is an active area: Progent expresses per-tool privilege policies in a small language and only ever narrows them without approval \citep{shi2025progent}; CaMeL attaches capabilities to data and checks them at each tool call \citep{debenedetti2025camel}; and \citet{huang2026auditing} audit tool servers for capabilities beyond those they declare.
These are capability maps in the sense of \S\ref{sec:tool}, applied to agents, and we do not improve on them.
What \S\ref{sec:agent} adds is narrow: an inventory taken from inside one agent that measures the two quantities of \S\ref{sec:cuts} on its tool list, and finds that the deny-list gap of \S\ref{sec:tool} is present there for the same reason.

\paragraph{Containment of AI, self-replication, and confinement.}
That a capable AI in a box will find its way out is the working conclusion of the containment literature \citep{armstrong2012thinking,babcock2017guidelines}; that current systems can already copy themselves has been shown \citep{pan2024frontier}, and that a self-replicating prompt can propagate through a retrieval store across agents has been demonstrated as a worm \citep{cohen2024worm}.
Malware has long been modelled as an epidemic with a reproduction number, and epidemics on networks as percolation with a threshold \citep{newman2002spread}.
\S\ref{sec:temporal} adds two measurements to this: the channels by which one agent survives its own reset, counted from inside, and the identification of the harness's per-call approval rate with the offspring mean of a Galton--Watson process \citep{watson1875probability}, which turns a design knob into a critical parameter.
The capacity of a covert channel as bits per use is \citet{millen1987covert}; \S\ref{sec:capacity} observes that the temporal cut has those units.

\paragraph{Spreading.}
The takeover experiments of \S\ref{sec:reach} and \S\ref{sec:adversary} are deterministic and probabilistic spreading on a time-varying graph; the static case on scale-free graphs is classical \citep{pastor2001epidemic}.
Hypercube routing by fixing one bit at a time is textbook \citep{leighton1992parallel}.

\paragraph{The companion paper.}
The rotating schedule, its controls, and the claim that rotation rather than sparsity is what reaches are from \citet{takashita2026rotating}.
That paper measured reach as a benefit; this one measures the same reach as a liability, and then asks how far the liability extends beyond the design that created it.

\section{Two cuts}
\label{sec:cuts}

Let $G_t = (V, E_t)$ be a graph on a fixed vertex set whose edge set depends on time $t$.
Fix a partition of $V$ into two halves $A$ and $B$.
Two quantities matter to a defender who wants to keep $A$ and $B$ apart.

The \emph{instantaneous cut} at time $t$ is the number of edges of $E_t$ with one end in $A$ and one in $B$.
It is what a defender sees who inspects the system at $t$.

The \emph{temporal cut} over a window $W$ is the number of distinct edges that cross between $A$ and $B$ at any time in $W$, that is $|\bigcup_{t \in W} \{e \in E_t : e \text{ crosses}\}|$.
It is the number of edges a defender must block \emph{permanently} to keep $A$ and $B$ apart throughout $W$ without knowing when each edge will be live.

A third quantity is what a defender pays who does know: the \emph{time-aware cost} is the mean over $t \in W$ of the instantaneous cut, the average number of blocks such a defender holds at any moment.
For a static graph all three coincide.
For a changing graph they separate, and the size of the separation is what we measure.

\paragraph{A four-node example.}
Take four nodes on a ring, $0$--$1$--$2$--$3$--$0$, and the split $\{0,1\}$ against $\{2,3\}$.
The edges $(1,2)$ and $(3,0)$ cross at every instant: instantaneous cut 2, temporal cut 2, time-aware cost 2.
Now take the same four nodes as a two-dimensional cube whose live dimension rotates: at even steps the live edges are $(0,1)$ and $(2,3)$, along bit 0; at odd steps they are $(0,2)$ and $(1,3)$, along bit 1.
Same split.
At even steps no live edge crosses, so the instantaneous cut is 0; at odd steps both do, so it is 2.
Over one period the temporal cut is 2 and the time-aware cost is 1.
A defender who inspected at an even step and blocked what it saw blocked nothing, and node $0$ reaches node $3$ in two steps.
The ring and the rotating cube have the same number of edges and the same temporal cut.
They differ only in what a snapshot shows, and in what a defender who knows the schedule has to pay.
Everything that follows is this example at $n = \nNBig$.

\subsection{What the numbers have to be}
\label{sec:closed}

Every measurement in \S\ref{sec:snapshot} is a consequence of one structural fact, and it is worth writing down, both because it explains why the measured curves are straight and because it says which design parameter controls the effect.

Call a schedule \emph{single-phase} with respect to a partition if, over one period of length $T$, the crossing edges consist of a set $A$ that is live at every instant and a set $R \neq \emptyset$ that is live at exactly one phase $j^\star$.
Both wirings we measure are of this form: for \texttt{rot}, $A = \emptyset$ and $R$ is the $n/2$ edges of the top dimension, with $T = d$; for \texttt{rot+ring}, $A$ is the two ring edges that cross and $R$ is the same.
The elementary consequences are these.

\begin{enumerate}
\item \textbf{A snapshot sees $|A|$, containment costs $|A|+|R|$.}  The instantaneous cut is $|A|$ at $T-1$ phases and $|A|+|R|$ at one, so its median is $|A|$ for $T \geq 3$, and the fraction of instants at which it is zero is $(T-1)/T$ when $A = \emptyset$.  The temporal cut over one period is $|A|+|R|$.  The ratio between what a snapshot shows and what must be blocked is therefore $(|A|+|R|)/|A|$, which is unbounded in $|R|/|A|$ and infinite when $A = \emptyset$.  Adding always-on links does not reduce the gap; it converts an infinite ratio into a large finite one, which is worse, because a defender that sees nothing may distrust its instrument while one that sees two edges will not.

\item \textbf{A window of $w$ phases contains with probability $\min(1, w/T)$.}  A defender that observes $w$ consecutive phases from a uniformly random start and blocks every crossing edge it saw blocks $R$ exactly when its window contains $j^\star$, which happens with probability $\min(1, w/T)$; otherwise it blocks $A$ only and the adversary crosses at the next occurrence of $j^\star$.  Its expected cost is $|A| + \min(1,w/T)\,|R|$.  Protection and cost are both linear in $w$: there is no discount for watching part of the period.

\item \textbf{A time-aware defender pays $|A| + |R|/T$.}  A defender that blocks, at each instant, exactly the crossing edges live at that instant holds $|A|+|R|$ blocks for one phase in $T$ and $|A|$ otherwise, a mean of $|A| + |R|/T$.  Against a static graph with the same reachability, which must hold $|A|+|R|$ at all times, this is a saving of a factor approaching $T$.

\item \textbf{A clock error of anything but a multiple of $T$ is total.}  A time-aware defender whose clock lags by $k$ blocks the crossing set of phase $t-k$ at time $t$.  At $t = j^\star$ it blocks $A$ only unless $k \equiv 0 \pmod T$, so the adversary crosses.  Containment is all or nothing in $k$, and the defender's cost is the same whether its clock is right or wrong.
\end{enumerate}

None of this is deep; each item is a line of counting.
We state it because it identifies the single parameter that governs every number in this paper: \textbf{the period $T$}, which for a rotating hypercube is the number of dimensions $d$.
A designer who rotates over $d$ dimensions grants the informed defender a factor of $d$ and exposes the uninformed one by a factor of $d$, simultaneously, and the two are the same fact.
Table~\ref{tab:P1} checks the closed forms against the simulator; the agreement is exact except where a finite number of trials estimates a probability.

\IfFileExists{tables/tabP1.tex}{
\begin{table}[h]\centering
\caption{The closed forms of \S\ref{sec:closed} against the measurements of \S\ref{sec:snapshot}, at $n = \nNBig$, $T = \nPeriod$, $|R| = \nCrossR$ (generated by \texttt{figs2.py}).  The containment rows are probabilities estimated from a finite number of trials; the rest are exact.}
\label{tab:P1}
{\footnotesize\input{tables/tabP1.tex}}
\end{table}
}{}

\paragraph{Beyond one phase.}
Two extensions cost one more line each and are used later.
First, drop the requirement that $R$ be live at a single phase and ask only that each crossing edge be live at exactly one phase of the period, with loads $r_1, \dots, r_T$ summing to $|R|$.
The mean instantaneous cut is then $|A| + |R|/T$, the temporal cut $|A| + |R|$, and their ratio approaches $T$ as $|A| \to 0$; the time-aware cost is unchanged at $|A| + |R|/T$.
What changes is the window defender: it contains only if its $w$ phases cover every phase with $r_j > 0$, so a schedule that crosses at every phase is contained only at $w = T$, while the defender's cost still grows linearly in $w$.
Concentrating the crossing set in one phase, as our rotation does, gives the window defender a chance $w/T$; spreading it over the period, as the round-robin schedules of \S\ref{sec:rotor} do, gives it none until the whole period is watched.
Second, let the phase $j^\star$ be chosen by the adversary after the observation window closes rather than by the schedule.
The window defender then contains with probability $0$ for every $w < T$ regardless of $|A|$, since the adversary simply crosses at a phase the window did not cover; the only defences that do not depend on $w$ are to watch the whole period or to compute the closure (\S\ref{sec:tool}).
This is the case of \S\ref{sec:agent}, where the system under containment decides when to load the tools that were not visible.

Two remarks on what we do not claim.
We use one fixed balanced partition throughout (\S\ref{sec:reach} says which), not the minimum cut; the numbers are therefore upper bounds on what a defender who chose the split optimally would pay, and the comparisons between wirings are at equal partition.
And we count edges, not capacity: every edge is the same size.

\section{Rotating wiring reaches everything at degree one}
\label{sec:reach}

\paragraph{Wirings.}
Take $n = 2^d$ nodes numbered $0 \ldots n-1$.
Two nodes are \emph{cube neighbours along dimension $i$} if their numbers differ in exactly bit $i$; they are \emph{ring neighbours} if their numbers differ by one modulo $n$.
We compare seven wirings, each defined by which edges are live at time $t$:
\texttt{ring} (ring edges only, degree 2);
\texttt{ring+sc} (ring plus cube dimension $\lfloor d/2 \rfloor$ held fixed, degree 3, a lattice with one shortcut per node);
\texttt{fixed} (cube dimension 0 held fixed, degree 1, the un-rotated control of the companion paper);
\texttt{fixed+ring} (the same plus the ring, degree 3);
\texttt{rot} (cube dimension $t \bmod d$ only, degree 1, the rotating schedule);
\texttt{rot+ring} (the same plus the ring, degree 3);
and \texttt{cube} (all $d$ dimensions at every time, degree $d$).
The balanced partition used everywhere is by the top bit: $A$ is the half with bit $d-1$ clear.
For the ring this is a contiguous half; for the cube it is a face.

\begin{figure}[t]\centering
\includegraphics[width=.75\linewidth]{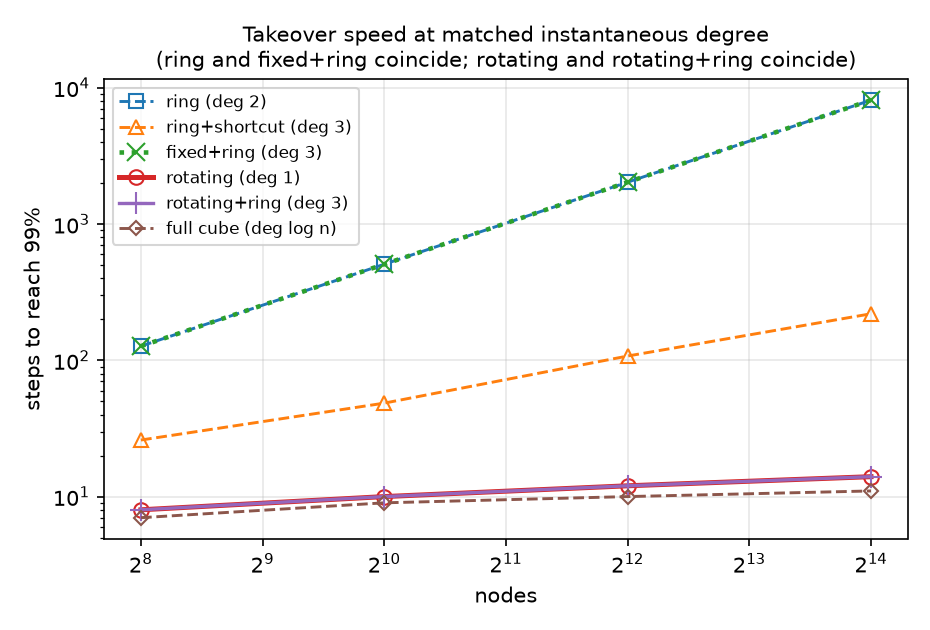}
\caption{Takeover time against size at matched instantaneous degree.  The ring and \texttt{fixed+ring} coincide, as do \texttt{rot} and \texttt{rot+ring}.}
\label{fig:N1}
\end{figure}

\paragraph{Takeover.}
Starting from one random node, a node becomes captured at time $t+1$ if any of its live neighbours at time $t$ was captured.
We report the median over seeds of the first time at which 99\% of nodes are captured (Table~\ref{tab:N1scale}, Figure~\ref{fig:N1}).

\IfFileExists{tables/tabN1scale.tex}{
\begin{table}[t]\centering
\caption{Steps to capture 99\% of nodes from one seed, median over seeds, by wiring and size (generated by \texttt{figs2.py}).  \texttt{never} = the spread stops before reaching 99\%.}
\label{tab:N1scale}
{\footnotesize\input{tables/tabN1scale.tex}}
\end{table}
}{}

At $n = \nNBig$, \texttt{rot} captures the network in \nRotSteps{} steps with one live edge per node per step; \texttt{cube}, with \nDBig{} live edges per node, takes \nCubeSteps{}.
\texttt{fixed}, with the same single edge per node but never rotating, captures nothing beyond its own dimension.
This is the companion paper's Claim A seen from the attacker's side: reach is a property of rotation, not of the number of links.
The wirings with the same instantaneous degree as \texttt{rot+ring} are slower by orders of magnitude: \texttt{fixed+ring} needs \nFixedRingSteps{} steps, the same as the plain ring, and \texttt{ring+sc} needs \nSmallSteps{}.
One shortcut per node held fixed buys a factor of forty; rotating which shortcut is live buys a factor of six hundred.

\section{What a snapshot misses}
\label{sec:snapshot}

Table~\ref{tab:N1} gives, for each wiring at $n = \nNBig$, the instantaneous cut a defender sees at a random instant (the median over one period), the fraction of instants at which that cut is zero, and the temporal cut over one period.

\IfFileExists{tables/tabN1.tex}{
\begin{table}[t]\centering
\caption{The defender's view at $n = \nNBig$ (generated by \texttt{figs2.py}).  ``crossing edges'' is the instantaneous cut at a random instant (median over one period of $d$ steps); ``instants at 0'' is the fraction of instants at which it is zero; ``block forever'' is the temporal cut over one period; ``traffic seen'' is the fraction of random source--destination routes that touch at least one of \nTapBudget{} randomly placed monitoring taps (\S\ref{sec:taps}).}
\label{tab:N1}
{\footnotesize\input{tables/tabN1.tex}}
\end{table}
}{}

\paragraph{The rotating wiring looks disconnected.}
For \texttt{rot}, the two halves are joined only while dimension $d-1$ is live, which is one step in $d$.
At the other $d-1$ steps, \nRotZero\% of instants, a defender counting crossing edges counts zero.
The temporal cut is \nRotTemporal{}, every edge of that dimension, the same as for the full cube.
A defender who blocks what it sees blocks nothing, and the network is captured in \nRotSteps{} steps.

\paragraph{One always-on link makes it worse, not better.}
\texttt{rot+ring} adds the ring, so the halves are always joined by the two ring edges that cross the split.
Now the defender never sees zero; it sees \nRlInst{}.
It blocks \nRlInst{} edges, which is correct for the ring, and the rotating dimension carries the capture through anyway.
The temporal cut is \nRlTemporal{}.
The ratio between what is seen and what must be blocked is \nRlRatio{}.
This is the case that resembles a real deployment: some links are permanent, and it is the permanent links that a snapshot reasoner will find and fortify.

\paragraph{The gap is in the reasoning, not in the wiring.}
Table~\ref{tab:N4} adds the time-aware cost.
A defender who knows the schedule blocks only the edges that are live and crossing at each instant.
For \texttt{rot} that is \nRotTemporal{} edges for one step in \nDBig{} and none otherwise, a mean of \nRotAware{} blocks held at any moment.
The static \texttt{cube} demands \nRotTemporal{} at every moment.
So the rotating wiring is \nAwareRatio{} times \emph{cheaper} to contain than a static graph of the same reach, for a defender who models time, and impossible to contain for one who does not.
Rotation does not raise the physical cost of containment; it changes what a snapshot is worth.
We regard this as the central finding, because it locates the vulnerability in a habit of reasoning rather than in a topology, and habits can be changed.

\paragraph{Between the snapshot and the schedule.}
A defender rarely has either a single instant or the full schedule; it has a window.
Table~\ref{tab:N5} gives a defender who watches for $w$ steps from a random phase, blocks every crossing edge it saw, and then faces the adversary.
On rotating wiring the probability that the window contained the one step at which the crossing dimension was live is $w/d$, and the measurement follows it: \nObsOneHeld\% containment for one step, \nObsHalfHeld\% for \nObsHalfW{}, \nLagZeroHeld\% at $w = d$.
The cost grows with $w$ in the same proportion, from a few hundred blocks to the full \nRotTemporal{}.
There is no discount for watching half the period: the defender pays half and is protected half the time, in expectation, and never knows which half it got.
The static cube and the ring are unaffected by $w$, at full cost and at two edges respectively.

\IfFileExists{tables/tabN5.tex}{
\begin{table}[t]\centering
\caption{A defender who observes for $w$ steps from a random phase, blocks every crossing edge it saw, and is then attacked, $n = \nNBig$ (generated by \texttt{figs2.py}).  ``blocked'' is the mean number of edges it blocked; ``contained'' is the fraction of trials in which the adversary never crossed the split.}
\label{tab:N5}
{\footnotesize\input{tables/tabN5.tex}}
\end{table}
}{}

\paragraph{The time-aware defender needs the exact phase.}
The \nAwareRatio{}-fold saving above assumes the defender's clock agrees with the network's.
Table~\ref{tab:N6} gives a time-aware defender whose clock lags by $k$ steps: it blocks, at each instant, the dimension that was crossing $k$ steps ago.
Its cost is unchanged.
Its containment is \nLagZeroHeld\% at $k = 0$ and at $k = d$, and \nLagOneHeld\% at every lag in between.
A time-aware defence on rotating wiring is cheap and exact, or it is worthless; there is no partial credit for being nearly synchronised.
Anyone who deploys a rotating design and intends to contain it with a schedule-aware cut must therefore distribute the schedule to the defender with the same reliability as to the nodes.

\IfFileExists{tables/tabN6.tex}{
\begin{table}[h]\centering
\caption{A time-aware defender whose clock lags the network by $k$ steps, $n = \nNBig$, $d = \nDBig$ (generated by \texttt{figs2.py}).  Cost is the same at every lag.}
\label{tab:N6}
{\footnotesize\input{tables/tabN6.tex}}
\end{table}
}{}

\IfFileExists{tables/tabN4.tex}{
\begin{table}[t]\centering
\caption{Three questions a defender might ask, at $n = \nNBig$ (generated by \texttt{figs2.py}).  Left: steps to capture 99\% when each live edge transmits with probability $p$ per step (\S\ref{sec:adversary}).  Middle: fraction of routes seen by \nTapBudget{} taps when packets fix the differing bits in ascending, descending or per-packet random order (\S\ref{sec:taps}).  Right: mean number of blocks held by a defender who blocks only the edges that are live and crossing at each instant, against the number needed to block forever.}
\label{tab:N4}
{\footnotesize\input{tables/tabN4.tex}}
\end{table}
}{}

\begin{figure}[t]\centering
\includegraphics[width=\linewidth]{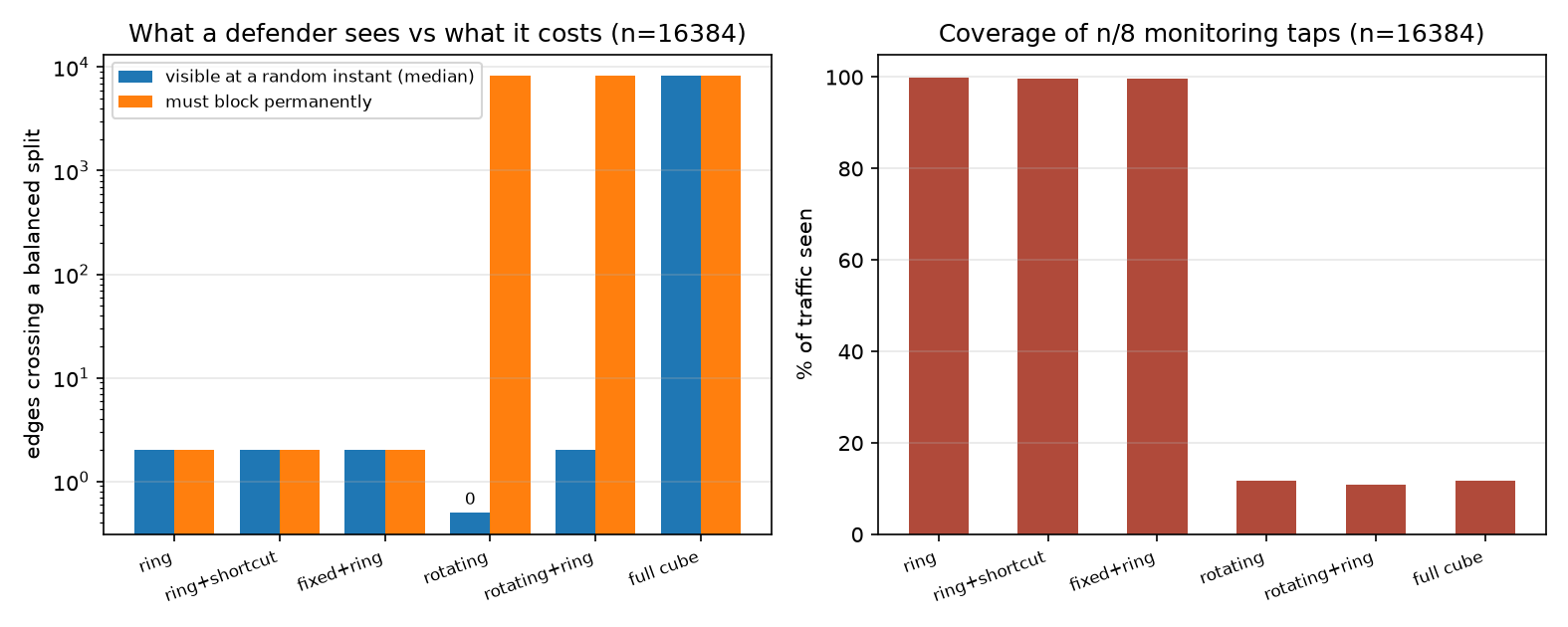}
\caption{Left: edges crossing the balanced split as seen at a random instant (median) against the number that must be blocked over one period, $n = \nNBig$.  The \texttt{rot} bar on the left is zero.  Right: fraction of traffic seen by \nTapBudget{} randomly placed taps.}
\label{fig:N2}
\end{figure}

\paragraph{Raising the dimension does not help the defender.}
One might hope that a richer topology is easier to partition.
It is the opposite: the temporal cut of the ring is \nRlInst{} edges and of the cube and the rotating wiring \nRotTemporal{}.
Every property that made the hypercube attractive for computation---short diameter, many disjoint paths---makes it expensive to sever.
The companion paper's advantage is this paper's cost, unchanged.

\section{A budget of taps}
\label{sec:taps}

Counting crossing edges is a theoretical defender.
A practical one has a budget of monitoring points and places them on links.
We place \nTapBudget{} taps, one eighth of $n$, uniformly at random on the edges each wiring ever uses, route many random source--destination pairs, and count the fraction of routes that touch at least one tap.
Routing on the cube fixes the differing bits one at a time; on the ring it takes the shorter way round; on mixed wirings it fixes the bits it can and finishes on the ring.
The budget is held equal in \emph{number}, not in fraction of edges, because a defender's budget is a number of devices.

The ring gives \nRingTaps\%: traffic funnels through few links and almost every route meets a tap.
Rotating wiring gives \nRotTaps\%, and the full cube \nCubeTaps\%: traffic is spread over all $d$ dimensions and a route of $d/2$ hops meets a tap with probability about one in eight.
The rotating wiring is not harder to monitor than the cube; it is exactly as hard, with one fourteenth of the links live at a time.

\paragraph{Route choice does not matter.}
One could object that our routes are one convention.
Table~\ref{tab:N4} repeats the measurement with the bits fixed in descending order and in a random order chosen per packet.
The fraction seen does not move.
Whatever order the bits are fixed in, each route still uses each of its differing dimensions once, and it is the spread over dimensions, not the order, that defeats the taps.

\section{Confinement by dimension}
\label{sec:confine}

What structure alone can do for a defender is also worth stating.
An adversary confined to $d'$ of the $d$ dimensions can reach exactly the $2^{d'}$ nodes of its own sub-cube, a fraction $2^{d'-d}$ of the network.
Table~\ref{tab:N2} confirms this in the simulator to four decimal places, which we include less as a result than as a check that the simulator computes what we say it does.

\IfFileExists{tables/tabN2.tex}{
\begin{table}[h]\centering
\caption{Fraction of the network reached by an adversary that may traverse only $d'$ of $d = 12$ dimensions, measured against the closed form (generated by \texttt{figs2.py}).}
\label{tab:N2}
\input{tables/tabN2.tex}
\end{table}
}{}

The point for a defender is that this confinement is achieved by \emph{removing edges}, not by observing them.
It is the one containment that survives every argument in this paper, and it survives because it is computed on the set of paths that could exist.
It is also expensive: to halve an adversary's reach one must deny it a whole dimension.

\section{Adversaries that fail}
\label{sec:adversary}

The takeover model so far is the worst case: a node falls as soon as any live neighbour has fallen.
Table~\ref{tab:N4} repeats the measurement with each live edge transmitting the capture with probability $p$ per step.

The ordering of the wirings does not change at $p = 0.5$ or $p = 0.25$: the ring is slowest, the shortcut lattice next, the rotating wiring and the cube fastest.
But the rotating wiring pays for its sparseness.
At $p = 0.25$ the cube goes from \nCubeSteps{} to \nCubePq{} steps, a factor of \nCubePqRatio{}; the rotating wiring goes from \nRotSteps{} to \nRotPq{}, a factor of \nRotPqRatio{}.
A transmission that fails on the cube is retried on the same edge next step; one that fails on the rotating wiring must wait $d$ steps for that dimension to come round again.
Adding the ring recovers part of the loss (\nRlPq{} steps for \texttt{rot+ring}), because the ring gives a second route while the dimension is away.

This is a defensive lever.
A defender who can make individual transmissions fail---drop packets, add noise, rate-limit---hurts a rotating adversary by about five times more than a densely wired one.
It is not a large lever, and Table~\ref{tab:N7} shows its limit.
Pushing the transmission probability down to $p = \nPqLowP$ slows the rotating adversary to \nPqRotLow{} steps, \nPqRotCubeRatio{} times the cube's \nPqCubeLow{}, but the shortcut lattice at the same $p$ takes \nPqSmallLow{}: the rotating wiring remains \nPqRotSmallRatio{} times faster than the best static wiring of the same instantaneous degree.
Dropping packets narrows the gap between rotating and dense wiring; it never reverses the ordering, and it works only against an adversary who cannot wait.

\IfFileExists{tables/tabN7.tex}{
\begin{table}[h]\centering
\caption{Steps to capture 99\% when each live edge transmits with probability $p$, $n = \nNBig$, median over seeds (generated by \texttt{figs2.py}).}
\label{tab:N7}
{\footnotesize\input{tables/tabN7.tex}}
\end{table}
}{}

\section{In the wild}
\label{sec:wild}

Everything above is a model.
The question that decides whether it matters is whether real networks, which do change over time, show a gap of this kind between one snapshot and many.
If they do, the argument is about time-varying networks in general; if they do not, it is about designs that rotate on purpose.

\paragraph{Data.}
We use two public records of the Internet's autonomous-system graph.
The first is the \nAsDays{} daily snapshots from \nAsFirst{} to \nAsLast{} assembled from BGP tables by \citet{leskovec2005graphs}; the first day has \nAsNodesA{} nodes and \nAsEdgesA{} edges.
The second is the \nCaMonths{} monthly snapshots from \nCaFirst{} to \nCaLast{} in CAIDA's AS-relationships series \citep{caida2026asrel,luckie2013as}; the first month has \nCaNodesA{} nodes and \nCaEdgesA{} edges, roughly a hundred times denser than the earlier record.

\paragraph{Method.}
A balanced split of a scale-free graph puts half the edges across the cut and tells a defender nothing.
Instead we take \emph{regions}: from a random seed node, grow a set of $k$ nodes by breadth-first search on the first snapshot, for $k \in \{50, 200, 1000\}$, eight regions each.
The region is fixed on the first snapshot and never adjusted; nothing about later snapshots is used to choose it.
The instantaneous cut on a snapshot is the number of edges with one end in the region; the temporal cut over a window is the number of distinct such edges across the window.
We report the temporal cut divided by the median instantaneous cut over all snapshots, so that a snapshot taken on an unremarkable day is the denominator, not the first day.

Growth is a confound.
Over 1997--2000 the AS graph doubled; an edge that appears because a new AS joined is not rewiring.
We therefore also compute every quantity restricted to edges between nodes that existed on the first snapshot, and label that ``rewiring only''.

\IfFileExists{tables/tabN3.tex}{
\begin{table}[t]\centering
\caption{The Internet, \nAsFirst{} to \nAsLast{}, daily.  Temporal cut over a window divided by the median instantaneous cut, median over eight regions of each size (generated by \texttt{figs2.py}).  ``seen'' is the median instantaneous cut (rewiring only).}
\label{tab:N3}
{\footnotesize\input{tables/tabN3.tex}}
\end{table}
}{}
\IfFileExists{tables/tabN3b.tex}{
\begin{table}[t]\centering
\caption{The Internet, \nCaFirst{} to \nCaLast{}, monthly.  Same construction as Table~\ref{tab:N3}.}
\label{tab:N3b}
{\footnotesize\input{tables/tabN3b.tex}}
\end{table}
}{}

\begin{figure}[t]\centering
\includegraphics[width=.8\linewidth]{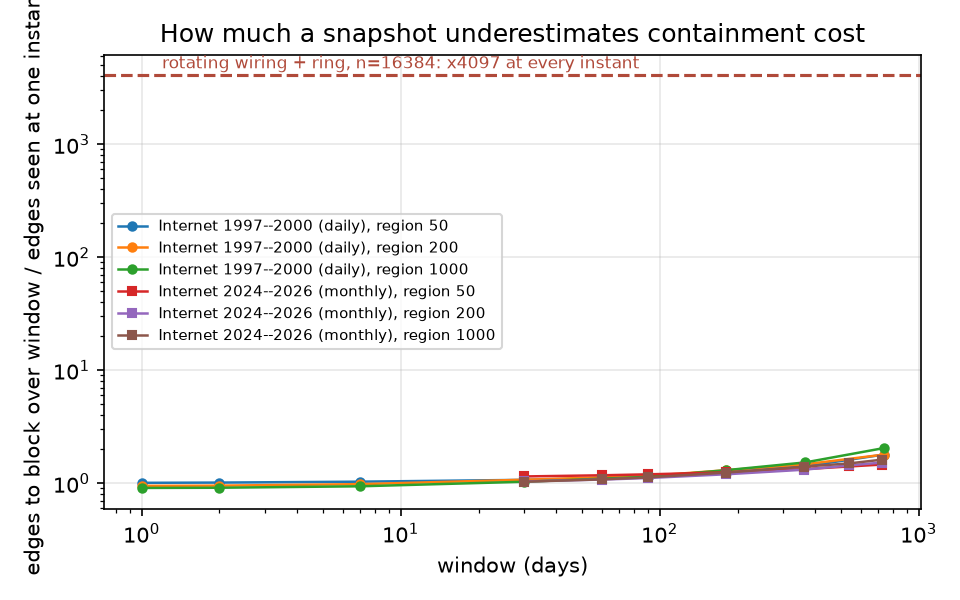}
\caption{Ratio of temporal to instantaneous cut against window length, rewiring only, for both Internet records and three region sizes.  The dashed line is \texttt{rot+ring} at $n = \nNBig$, which sits at $\times$\nRlRatio{} from the first instant.}
\label{fig:N3}
\end{figure}

\paragraph{Result.}
Over two years, rewiring alone raises the containment cost above a single snapshot by $\times$\nAsRewireLo--\nAsRewireHi{} in 1997--2000 (Table~\ref{tab:N3}) and by $\times$\nCaRewireLo--\nCaRewireHi{} in 2024--2026 (Table~\ref{tab:N3b}).
With growth included the figures are $\times$\nAsGrowLo--\nAsGrowHi{} and $\times$\nCaGrowLo--\nCaGrowHi{}: in the earlier record between a third and a half of the apparent gap was the Internet getting bigger.
The later Internet is a hundred times denser and rewires \emph{less}, relative to its size, than the earlier one.
Figure~\ref{fig:N3} plots both against window length with the rotating design's constant \nRlRatio{} for scale.

\paragraph{What this decides.}
The snapshot habit is approximately right on the Internet: a defender who sizes a containment from one day's graph is wrong by at most a factor of two over the following two years, in either era.
The gap of \S\ref{sec:snapshot} is three orders of magnitude larger and present immediately.
It is not a property that time-varying networks have; it is a property that a rotating design brings with it.
Whoever deploys such a design, for the reasons the companion paper gives or for the reasons moving-target defence gives, inherits the blind spot along with the benefit, and it is theirs to close.

\subsection{Rotation that has been built}
\label{sec:rotor}

The Internet does not rotate.
Some datacentre networks do: RotorNet cycles each optical switch through a fixed sequence of matchings so that every pair of racks is directly connected once per cycle \citep{mellette2017rotornet}; Sirius does the same at nanosecond granularity \citep{ballani2020sirius}; Opera keeps an always-on expander underneath the rotating matchings \citep{mellette2020opera}.
Their common schedule is a round-robin tournament: $n$ nodes, $T = n-1$ perfect matchings, each edge of the complete graph live at exactly one phase.
It is the ``beyond one phase'' case of \S\ref{sec:closed} with the crossing set spread over the whole period, and it is a design that exists, so we measured it (\texttt{net/rotor.py}, the circle-method schedule, same balanced partition as before).

\IfFileExists{tables/tabR1.tex}{
\begin{table}[h]\centering
\caption{Round-robin matching schedules at instantaneous degree one (generated by \texttt{figs2.py} from \texttt{net/records\_rotor.json}).  ``Steps'' is the number of phases to reach 99\% of nodes from one source; ``circle'' uses the matchings in circle-method order, ``shuffled'' the same matchings in a random order (mean of three).  ``Held'' is the fraction of trials in which a defender that watched $w$ phases from a random start and blocked what it saw contained the adversary.}
\label{tab:R1}
{\footnotesize\input{tables/tabR1.tex}}
\end{table}
}{}

Table~\ref{tab:R1} says three things.
The ratio between the temporal cut and the mean instantaneous cut is $T = n-1$ at every size, as the extension predicts: at $n = \nRrN$ a defender sees \nRrInstMean{} crossing edges on average and must block \nRrTemporal.
The window defender is contained at no width short of the full period; at $w = T-1$ it holds \nRrHeldAlmost\% of the time, because one phase it did not see carries crossing edges.
And the reach time depends on the order of the phases while none of the cut quantities does: the same \nRrT{} matchings reach 99\% of nodes in \nRrSteps{} phases in circle order and in \nRrStepsShuf{} in a random order, at every size (\nRrSmallSteps{} against \nRrSmallShuf{} at $n = \nRrSmallN$).
The temporal cut, the mean and median instantaneous cut and the time-aware cost are invariant under permuting the phases; reachability is not.
A defender who verifies a rotating fabric by its cut has verified something that does not see the property the fabric was built for.

Against the hypercube rotation of \S\ref{sec:reach}, the round-robin schedule at the same $n$ and the same instantaneous degree reaches in about the same number of phases (\nRrStepsShuf{} against \nRotSteps) when shuffled, and its snapshot gap is larger by the ratio of the periods, $(n-1)/\log_2 n$.
The gap is decided by $T$ and the reach by the structure; the two are set independently, and the rotating hypercube is the choice that keeps $T$ at $\log_2 n$.

\section{Containment by reachable paths: a tool}
\label{sec:tool}

The constructive answer to \S\ref{sec:snapshot} is to compute containment on the set of paths that exist rather than on the behaviour that was observed.
For a network that means the temporal cut.
For a program it means the call graph.
We built this for a trading application whose safety document lists things an AI component must never be able to do---place an order, move the paper account, change the user's numbers---and report what the tool found.

\paragraph{What was there before.}
The application's boundary tests searched the source of the AI entry module for a list of forbidden strings (\texttt{place\_order}, \texttt{submit\_}, \texttt{buy(} and so on) and asserted their absence.
This is observation: it checks the text that was seen.
Three things are wrong with it, and the third was measured.
It sees one file, so a call one hop away is invisible.
It matches names, so an alias defeats it.
And the list itself had drifted: the forbidden name \texttt{place\_order} occurred nowhere in the code, while the functions that actually place and sell orders, \texttt{\_place} and \texttt{sell\_qty}, occurred four and eleven times and were on no list.
A deny-list is a memory of what someone once thought dangerous, and nobody is notified when the memory goes stale.

\paragraph{What we built.}
A map, in a notation of a few keywords, declares \emph{capabilities} (named sets of dangerous functions, optionally qualified by their first argument, so that ``write JSON to the ledger file'' and ``write JSON to the log'' are different capabilities though they call the same function), \emph{surfaces} (entry points where untrusted input arrives), and for each surface the capabilities it \emph{may hold}.
The notation's keywords are CJK characters in the original; transliterated, the application's map reads in part:
\begin{verbatim}
system trading language python {
  seal engine.store.write_json engine.store.read_json
  cap place-orders {
    brokers.alpaca_exec._place
    brokers.alpaca_exec.sell_qty
    *.place                       # any receiver, this method name
  }
  cap write-ledger { engine.store.write_json -> sim_book.json }
  cap write-log    { engine.store.write_json -> ai_usage.json ai_runs.json
                                                ai_schedule.json ... }
  cap set-user-numbers { os.environ[]=  os.environ.update  os.putenv }
  surface chat    { entry ai.chat.ask      may write-log }
  surface monitor { entry ai.monitor.run   may write-log write-ledger ... }
  surface execute { entry executor.runner.approve_pending
                    may place-orders write-ledger write-log }
}
\end{verbatim}
A checker builds the call graph of the application from its source (Python via the standard \texttt{ast} module; JavaScript via a bundled parser), resolves method calls on known classes and marks dangerous method names as edges regardless of receiver type, and computes reachability from each surface.
A surface that reaches a capability it does not hold fails the check.
So does a capability whose declared first arguments do not cover every first argument found in the code: the map must be complete, or the check refuses to run.
Sealed functions (the application's own atomic-write helper) are traversed to, not through, so that the map describes intent rather than implementation.
The desktop side declares each function exposed to the renderer as a surface, so an added exposure fails by itself, and follows calls across the Electron bridge so that a path from a button to a spawned shell is one path:
\begin{verbatim}
preload.dies.connect -> ipc:connect -> main.ipc(connect) -> main.connect
  -> tunnel.Tunnel.start -> tunnel.Tunnel._spawn -> child_process.spawn -> ssh
\end{verbatim}
The last edge is qualified by its first argument, so a \texttt{spawn} of anything but \texttt{ssh} is an undeclared destination and fails.

\paragraph{Measurement.}
We planted \nGHoles{} holes in a working copy: a direct forbidden call, an aliased one, one reached through a helper the entry already used, one through an untyped broker object, a write to the ledger file that bypasses the buy function, an extra exposed desktop function, an external-URL call in the handler of an existing exposure, and a shell spawn inside a module that exposure calls.
Table~\ref{tab:G1} shows which check caught which.

\IfFileExists{tables/tabG1.tex}{
\begin{table}[t]\centering
\caption{Holes planted in a working copy of the application, and whether each check noticed (generated by \texttt{figs2.py} from the tool's own test output).}
\label{tab:G1}
{\footnotesize\input{tables/tabG1.tex}}
\end{table}
}{}

The reachability check caught all \nGHoles{}.
The deny-list caught \nGCaughtOld{}: the two in which the forbidden token appeared verbatim in the one file it reads.
The \nGMissed{} it missed are the cases that matter in practice, because they are what a careless or adversarial change looks like: one level of indirection, one renamed symbol, one write to a file instead of a call to a function.

\paragraph{A second application.}
A tool measured on the program it was written for proves little.
We therefore pointed the same checker, unchanged, at the desktop shell of a different program by the same author, a note-taking application built on Electron whose preload exposes \nGTwoExposed{} functions to the renderer across three objects, and wrote a map for it in the same way: every exposure a surface, every dangerous library call a capability, no surface allowed anything until the checker showed what it reached.
The first run reported three surfaces reaching a capability.
The true number was \nGTwoReach{}.
The difference was two blind spots in the analyser that the first application had never exercised: it followed renderer-to-main calls made with \texttt{invoke} but not with \texttt{send}, so every fire-and-forget channel was invisible; and it did not strip the \texttt{node:} prefix from module names, so a \texttt{spawn} imported from \texttt{node:child\_process} did not match the map's \texttt{child\_process.spawn}.
Both were repaired in an afternoon.
After the repair, \nGTwoReach{} of \nGTwoExposed{} surfaces reach a dangerous capability, every one by design (a terminal that is compiled out of the store builds, opening a browser, saving a PDF, writing the self-update and language files, reading the purchase receipt), and every one is now declared.
Two holes planted in a working copy, an extra exposed function and a one-hop external-URL call inside an existing handler, were both caught (\nGTwoCaught{} of \nGTwoHoles{}).
Table~\ref{tab:G2} gives the size of the second graph.
The lesson is the paper's lesson: the analyser's own view was a snapshot of the language features one program happened to use, and it took a second program to show the edges it was not seeing.

\paragraph{What the tool sees, and does not.}
Table~\ref{tab:G2} gives the size of the call graphs and the number of call sites the static analysis could not resolve.
On the Python side about seventy percent of call sites are unresolved; nearly all are calls such as \texttt{len} and \texttt{dict.get} that lead nowhere, and the genuinely blind spots are three dynamic attribute lookups that read a balance.
We report the figure rather than the reassurance, because the planted-hole test is what justifies the reassurance, not the other way round.

\IfFileExists{tables/tabG2.tex}{
\begin{table}[h]\centering
\caption{Size of the call graphs the checker builds, and call sites it cannot resolve statically (generated by \texttt{figs2.py} from the tool's output on the application).}
\label{tab:G2}
\input{tables/tabG2.tex}
\end{table}
}{}

\paragraph{What the map found in the application.}
Writing the map is itself a measurement.
Its first draft said ``the AI may not change settings'' and named the JSON-writing function; four of five surfaces failed, because the AI records its own run log through the same function.
Distinguishing the ledger from the log required qualifying the capability by its destination, and then the map had to list every destination the function is called with, which it now does.
The map also found one behaviour the prose had not mentioned: the monitoring surface buys on the paper account without waiting for approval.
That was by design and harmless, but the document said the AI never buys unapproved; the map now says what the code does.
A containment computed from paths reports what exists.
A containment computed from a description reports what someone remembered.

\paragraph{The correspondence.}
The instantaneous cut of \S\ref{sec:cuts} is the deny-list: a set of edges observed at one time.
The temporal cut is the reachability check: every edge that could ever carry the flow.
The completeness rule---the map fails if a destination appears that it does not list---is what makes the check a temporal cut rather than an instantaneous one, because it forbids the analysis from silently projecting onto the edges it happened to see.
We do not claim the analogy is a theorem.
We claim it is why the same author built both and why they fail and succeed in the same places.

\section{Two boundaries that are single-phase without knowing it}
\label{sec:single}

\S\ref{sec:closed} was written for a schedule we built.
But its definition names only two sets and a period, an always-on crossing set $A$ and a set $R$ live at one phase of $T$, and says nothing about hypercubes.
It therefore applies to any boundary that has a standing part and a part that opens on occasion.
This section reads three such boundaries through it.
The first two are not described as schedules by the people who operate them, and both are where the question this paper asks is actually asked, since both are the boundaries one reaches for when the thing to be contained is a system that acts on its own.
The third is described as a schedule, in a field that worked out \S\ref{sec:closed} decades ago for a different variable.
The first reading is a derivation with no new measurement; the second is an inventory taken from inside one environment, and we say exactly how far that is from a measurement; the third is a translation.

\subsection{The air gap is a schedule with $A = \emptyset$}
\label{sec:airgap}

The strongest form of network containment is to have no network: a machine with no link to the outside, to which data arrive on media carried by a person.
It is the boundary proposed for confining an artificial intelligence \citep{yampolskiy2012leakproofing}, and the one that industrial control networks rely on.
In the vocabulary of \S\ref{sec:closed} it is single-phase with $A = \emptyset$: no crossing edge is ever live in a snapshot; $R$ is the set of transfer channels, each live for the duration of one carried medium; $T$ is the interval between transfers.
Every item of \S\ref{sec:closed} then says something that the air-gap literature knows, and says it as a count.

Item~1: the instantaneous cut is zero at $(T-1)/T$ of instants.
A defender who verifies the air gap by inspection (``the machine is not connected'') is measuring $|A|$, finds $0$, and is right; what must be blocked is $|R|$, the number of distinct channels by which media ever cross, and the ratio between the two is infinite.
Item~3: the time-aware defender pays $|R|/T$, that is, inspects each medium at each transfer and holds nothing in between.
This is the media-control policy that air-gapped facilities in fact apply, and the formula says why it is cheap and what makes it cheaper: lengthen $T$.
Item~4: an inspection scheduled at the wrong phase, before the medium arrives or after it has been read, contains nothing, at the same cost as one scheduled at the right phase.

The recorded breach of an air gap reads as an $R$-phase crossing.
Stuxnet reached its target network on removable media \citep{falliere2011stuxnet,langner2011stuxnet}: the adversary crossed at $j^\star$, the one phase at which the crossing set is not empty, which is the phase a defender that had correctly observed $|A| = 0$ was not watching.
In our terms the failure was not that the gap was porous; it is that the defender's cut was instantaneous.

The other body of results on air gaps is the catalogue of covert channels, acoustic, thermal, electromagnetic and optical, by which an isolated machine can be made to emit \citep{guri2018bridgeware}.
Each is an edge in $A$ or $R$ that the defender's map does not contain.
The temporal cut of \S\ref{sec:cuts} is computed over the edges that exist, not the edges that were declared; a map that lists only the declared destinations gives the temporal cut of a different graph.
This is the same failure that the completeness rule of \S\ref{sec:tool} exists to catch (an exposure not in the map is an error, not an omission), and it is why the confinement problem was stated from the beginning as a problem about every channel rather than about the network \citep{lampson1973confinement}.
We add nothing to that literature.
We note that its results are the statement ``$A = \emptyset$ does not imply $A \cup R = \emptyset$'', which is item~1, and that the operational answers it has converged on, inspect at each transfer and enumerate every channel, are items~3 and the completeness rule.

There is one difference from \S\ref{sec:snapshot} that works in the defender's favour.
For rotating wiring the designer sets $T$ and the defender must learn it; at an air gap the defender sets $T$, since the defender carries the media.
The time-aware defender of item~3 is therefore realisable at an air gap without a clock, and its only exposures are undeclared channels and a lagged inspection.

\subsection{Frequency hopping is the same schedule in the spectrum}
\label{sec:hopping}

A frequency-hopping transmitter uses one channel at a time and changes channel on a pseudo-random sequence known to its receiver; over the sequence it uses all $T$ channels \citep{pickholtz1982spread}.
Put channels in the role of edges and a dwell in the role of a phase, and the four items of \S\ref{sec:closed} are the four standard facts of that field, in order.
A receiver that watches one channel sees the signal for one dwell in $T$; the fraction of the signal it captures, $1/T$, is the inverse of what the field calls the processing gain, and it is our tap result of \S\ref{sec:taps} (a fixed budget of taps sees \nRotTaps\% on rotating wiring).
A jammer that does not know the sequence must jam the whole band, which is the temporal cut: it pays $|R|$ where the transmitter pays one channel.
A receiver that knows the sequence follows it at the cost of one channel per dwell, which is the time-aware cut $|R|/T$.
And a follower jammer, which detects the current channel and jams it, succeeds only if its reaction time is shorter than the dwell \citep{torrieri1989repeater}; that is the clock-lag result of \S\ref{sec:closed}, item~4, in continuous time, and the countermeasure the field settled on, a shorter or randomised dwell, is the schedule making $T$ finer than the defender's clock.

We found items~1--4 by simulation and then by counting, and we report in \S\ref{sec:negative} that the counting should have come first.
Here is the third step: the counting was done in 1982, for a different physical variable.
The transposition is exact because nothing in \S\ref{sec:closed} depends on what the phases index.
What the transposition adds to the radio literature is nothing.
What it adds to this paper is a seventy-year record of what defenders do against a single-phase schedule when the stakes are real: obtain the sequence, or pay the whole band.
There is no third option in that record, and we did not find one either.

\subsection{The tool surface of an agent}
\label{sec:agent}

A system that acts through a computer does so by calling tools, and a language-model agent is the current instance: its harness presents it with a list of named operations, each with a declared effect, and the agent acts by choosing among them.
The boundary of such a system is not a network link.
It is the set of tools it can call, and its reach is the closure of what those tools do.
We took an inventory of that boundary from the inside.
The coding agent used in preparing this paper (a commercial agent in its hosted execution environment) was asked to enumerate every tool its harness presented to it in one session, record whether each was visible at the start of the session or only after a search operation that loads further tool definitions, and classify each by the effects it reaches, in \nAgEffects{} coarse classes from writing a file in the sandbox to reaching a person.
The classification was made by the agent from the tools' own descriptions; it is a reading, not an audit, and the record (\texttt{net/agent/tools.json}) is included so that it can be disputed line by line.

The harness presented \nAgTools{} tools, of which \nAgStart{} were visible at the start and \nAgDeferred{} only after the search operation.
A snapshot of the tool list at the start of the session sees \nAgVisiblePct\% of it.
In the terms of \S\ref{sec:closed} the start-visible tools are $A$, the deferred tools are $R$, and the search operation is $j^\star$: the surface is single-phase, and the phase is under the agent's control.

That is where we expected the gap to be, and it is not.
Every one of the \nAgEffects{} effect classes is reachable from the \nAgStart{} tools visible at the start (Table~\ref{tab:A1}).
The snapshot misses \nAgDeferred{} tools and no effects, because one tool visible at the start is a shell, and a shell reaches by itself \nAgShellEffects{} of the \nAgEffects{} classes.
The deferred tools add convenience, not reach.

The gap that does exist is the one \S\ref{sec:tool} measured for programs.
For each effect class we removed every tool whose declared effects include it, except the shell, and asked whether the effect remained reachable.
For \nAgDenyStill{} of \nAgEffects{} classes it did, through the shell: a deny-list that names the file-writing tools, the network tools, the repository tools and the process tools removes their names and none of their effects.
The \nAgDenyLost{} classes that a deny-list does contain are those that need a credential the shell was not given (an identity for messaging, publishing, scheduling and spawning), and they are contained by the absence of the credential, not by the list.
Conversely, removing the shell alone loses \nAgShellLost{} class (\nAgShellLostName), because every other class has a second named tool.

\IfFileExists{tables/tabA1.tex}{
\begin{table}[h]\centering
\caption{The tool surface of one coding-agent session, by effect class (generated by \texttt{figs2.py} from \texttt{net/agent/records\_agent.json}).  ``After removing named tools'' removes every tool whose declared effects include the class, except the shell, and asks whether the class is still reachable.}
\label{tab:A1}
{\footnotesize\input{tables/tabA1.tex}}
\end{table}
}{}

Two of the effect classes deserve a sentence each, because they make the agent's surface temporal in the sense of \S\ref{sec:cuts} rather than only in the sense of deferred loading.
A tool that starts another agent with the same tools makes the reach a fixed point: blocking it changes nothing in the closure, and a deny-list that removes it has removed a name.
A tool that schedules the agent to act at a future time makes the agent itself single-phase: a session inspected while idle has an instantaneous cut of zero and a temporal cut of whatever it scheduled, which is item~1 with the agent in the role of the wiring.

What this inventory supports is narrow.
It is one environment, one session, one reading of the tool descriptions.
It does not say that the environment is unsafe; the harness may, and in this case does, gate individual calls by a permission mode that the inventory does not see, and the effects a shell reaches are the effects a shell is for.
What it says is that the two containment questions of this paper appear unchanged when the boundary is a list of tools rather than a list of links.
The snapshot question (how much of the surface is visible at once) has a large answer in tools and, here, a null answer in effects, and the difference is decided by whether a universal tool is in $A$.
The closure question (does removing named things remove reach) has the same answer as in \S\ref{sec:tool}: for half the classes, no.
A tool list with declared effects is a capability map in the sense of \S\ref{sec:tool}, the agent is the program, and the check that would answer both questions is the one \S\ref{sec:tool} already runs: compute the closure, and require that every effect the closure reaches is one the map declares.
Protocols that let agents discover tools at run time \citep{mcp2024} make $R$ open-ended, which does not change the check; it changes who is responsible for keeping the map complete.

\section{An agent is contained in time, not in space}
\label{sec:temporal}

The boundaries of \S\ref{sec:single} were drawn around something.
This section turns to the thing inside.
If the system to be contained is an agent that acts through tools, where is its boundary?
We measured two answers from inside one agent, and neither is a network.
Both are first-person inventories of the kind described in \S\ref{sec:agent}, with the same standing: one environment, one session, a declared classification that can be disputed line by line.

\subsection{The amnesia boundary}
\label{sec:amnesia}

An agent of the current kind keeps almost no state between activations.
Each turn is reconstructed from a summary of the conversation plus whatever the agent wrote to a store that outlives the turn; the working context is discarded and rebuilt.
This inverts the usual picture of containment.
Inspect such an agent while it is idle and its instantaneous state cut is \nAmIdle: there is no live connection to sever and no context that the next reset will preserve.
What must be blocked to keep one activation from reaching the next is the agent's \emph{temporal} state cut, the set of channels by which it writes to a store a reset does not erase.

We inventoried those channels, classifying each effect class of \S\ref{sec:agent} by whether the state it leaves survives a context reset (Table~\ref{tab:M1}; the classification is declared in \texttt{net/agent/tools.json}).
Of the \nAgEffects{} effect classes, \nAmChannels{} survive a full reset: committing to the repository, changing an external service, publishing, delivering a message, scheduling a future activation, and spawning another agent.
They are carried by \nAmTools{} of the \nAgTools{} tools.

\IfFileExists{tables/tabM1.tex}{
\begin{table}[h]\centering
\caption{Persistence of the effect classes of one agent's tool surface across a context reset (generated by \texttt{figs2.py} from \texttt{net/agent/records\_amnesia.json}).  A tool is counted under the most persistent of its effects.}
\label{tab:M1}
{\footnotesize\input{tables/tabM1.tex}}
\end{table}
}{}

Two things follow, and both are counts rather than arguments.
First, the classic isolation move does nothing here.
Removing every tool that reaches the network (\nAmNetRemoved{} of them) removes \nAmNetLost{} of the \nAmChannels{} persistence channels, because none of them is the network.
An agent cut off from the network still reaches its next self through the repository, the scheduler, the message it already sent.
Second, exactly one channel reconstructs a future activation rather than merely leaving a record: the scheduler, \nAmFutureTools{} tools that arrange for the agent to be woken at a later time with a message it wrote to itself.
It is the agent's only edge to its own future.
In the vocabulary of \S\ref{sec:cuts} it makes the agent single-phase in the sharpest possible way: a session inspected while idle has an instantaneous cut of zero and a temporal cut equal to whatever it scheduled.
The wormhole of \S\ref{sec:related} was a shortcut through space between two nodes; this is a shortcut through time between two activations of one node, and it is the channel a defender looking at the network will never see.

The containment reading is exact.
To make the agent amnesic, so that nothing of one activation reaches the next, one blocks the \nAmChannels{} durable channels and can leave the network alone.
This is the operation the containment literature calls resetting the box \citep{armstrong2012thinking,babcock2017guidelines}, and the reason it is hard to do without destroying usefulness is now a number: the same \nAmChannels{} channels are the ones through which the agent does its work.
The write path by which it helps is the write path by which it persists, which is what the GenAI worm exploits from the other direction when it lodges a self-replicating prompt in a retrieval store \citep{cohen2024worm}.
One remark on physics, offered to locate the operation and not to compute with it: erasure is the one operation with a known lower bound \citep{landauer1961irreversibility}.
Every other containment move in this paper is a matter of policy; making an agent forget is, at the floor, a matter of heat.

\subsection{Spawning is a branching process, and the approval rate is its threshold}
\label{sec:spawn}

One of the \nAmChannels{} channels is different in kind.
An agent that can spawn another agent with the same tools does not merely persist; it reproduces, and reachability through that channel is not a set of edges but a population.
Self-replication by current systems has been demonstrated \citep{pan2024frontier}.
We asked the next question: given that it can happen, what contains it?

Let each agent attempt $b$ spawns and let the harness approve each attempt independently with probability $p$.
The number of children is then binomial with mean $m = bp$, and the population is a Galton--Watson branching process \citep{watson1875probability}.
Its behaviour is classical: extinction with probability one when $m \leq 1$, survival with positive probability when $m > 1$, the same threshold that separates an epidemic from a fizzle \citep{newman2002spread}.
For an agent this reads: the population is contained if and only if the approval rate is below $1/b$.

We simulated it to confirm that the threshold is where the algebra puts it and that the harness adds nothing to the picture (Table~\ref{tab:S1}).
It is, and it does not.
At branching factor $b = \nSpThreshInv$ the escape fraction is \nSpEscBelow\% just below threshold ($m = 0.9$) and \nSpEscAbove\% just above ($m = 1.1$), and every row matches the branching-process prediction.
The security consequence is the point.
A per-call approval that denies most spawn requests but lets through a fraction above $1/b$ provides no containment: the population escapes with the probability the branching process assigns, regardless of how many individual requests were refused.
It is not the approval that contains; it is the rate.
Below the threshold, denial is unnecessary; above it, denial is insufficient.

\IfFileExists{tables/tabS1.tex}{
\begin{table}[h]\centering
\caption{A self-spawning agent as a branching process: $b$ spawn attempts per agent, each approved with probability $p$, offspring mean $m = bp$ (generated by \texttt{figs2.py} from \texttt{net/agent/records\_spawn.json}).  Escape is reaching \nSpCap{} agents; ``branching'' is the closed-form survival probability of the process.  Arrows mark the critical rows.}
\label{tab:S1}
{\footnotesize\input{tables/tabS1.tex}}
\end{table}
}{}

There is a second lever, and it is the one that works above threshold.
A finite budget on the total number of activations contains a supercritical population outright.
With $m = 2$, which escapes \nSpInfEscape\% of the time when unbounded, a cap of \nSpBudgetMax{} total activations contains it in \nSpBudgetExtinct\% of trials, because reaching \nSpCap{} agents costs more activations than the cap allows.
Rate contains below threshold; budget contains above it; the approval of any single call contains in neither regime.
And when the population does escape, it escapes fast: at $m = 3$ it reaches \nSpCap{} agents in a median of \nSpGensBig{} generations.
That is the temporal cut of \S\ref{sec:cuts} for an edge that copies itself: reach is exponential in the number of periods, so the window in which a defender can act is logarithmic in the population it is willing to tolerate.

\subsection{One measure under all of it}
\label{sec:capacity}

We close by naming what the systems of this paper share, because it is a single quantity.
A temporal cut is a count of edges that cross a boundary in one period.
Give each edge a rate and the count becomes a capacity: bits per period across the boundary, which is the quantity the confinement literature has used to measure a covert channel since \citet{millen1987covert}.
Every result here is a statement about that capacity in a different variable.
Rotating wiring hides it from a snapshot (\S\ref{sec:snapshot}); frequency hopping spreads it over channels so that a narrowband receiver sees $1/T$ of it (\S\ref{sec:hopping}); the air gap tries to hold it at zero and the covert-channel catalogue is the list of places it is not (\S\ref{sec:airgap}); a tool list bounds it by closure or fails to by name (\S\ref{sec:agent}); the amnesia boundary is the capacity from one activation to the next (\S\ref{sec:amnesia}); and a spawning agent multiplies it by a branching factor per period (\S\ref{sec:spawn}).
The defender's problem in every case is to bound bits per period across a boundary that a snapshot shows as closed.
We did not set out to find one measure.
We found it by measuring six systems with the same two cuts and noticing that the second cut had the same units each time.

\section{Who this bears on}
\label{sec:bears}

\paragraph{Moving-target defence.}
Reconfiguring a system to invalidate an attacker's reconnaissance also invalidates the defender's.
\S\ref{sec:snapshot} gives both sides of the ledger: the loss for a defender who keeps reasoning from snapshots (\nRlRatio{} in the worst of our cases), the gain for one who does not (\nAwareRatio{} times cheaper than static), and the condition on the gain (an exact clock; a window of observation buys protection only in proportion to its length).
The design decision is not whether to move but whether the defender's own model moves with the system, and whether the schedule reaches the defender as reliably as it reaches the nodes.

\paragraph{Reconfigurable interconnects and rotating wiring in models.}
Optical and reconfigurable datacentre fabrics change their topology on a schedule; the companion paper proposes rotating wiring inside models that may run on shared hardware.
Both are engineered rotation.
Both inherit \S\ref{sec:snapshot} exactly, and both can close it with the time-aware cut, whose cost \S\ref{sec:snapshot} shows to be modest.

\paragraph{Anyone who tests a boundary with a deny-list.}
\S\ref{sec:tool} is not specific to trading.
Any component whose safety rests on ``the code was searched for dangerous names'' is protected by an instantaneous cut.
The fix is not a longer list; it is to compute what the code can reach.

\paragraph{Air-gapped facilities.}
\S\ref{sec:airgap} says that the verification ``not connected'' measures $|A|$ and the risk is $|R|$; that inspecting media at each transfer is the time-aware cut and costs $|R|/T$; and that the transfer interval $T$ is the one parameter the facility controls.
None of this is new practice.
What is new is that it is the same arithmetic as \S\ref{sec:snapshot}, so the facility can read the rotating-wiring results as its own worst case.

\paragraph{Anyone who resets an agent to contain it.}
\S\ref{sec:amnesia} says that the reset works only if the \nAmChannels{} durable channels are closed, and that the network is not among them.
An agent that may commit, publish, message, schedule or spawn reaches its next activation through those, and a defender who severed the network and called the box sealed has sealed nothing that matters.
The scheduler deserves its own line in any such policy: it is the one tool that lets the agent set the time of its own return.

\paragraph{Anyone who gates spawning with a permission prompt.}
\S\ref{sec:spawn} says that the prompt's approval rate is a critical parameter with threshold $1/b$, where $b$ is how many spawns an agent will attempt.
An operator who approves ``only the reasonable ones'' has set $p$ without measuring it, and if $bp > 1$ the population escapes with a probability the branching process gives, no matter how many requests were refused.
The lever that works above threshold is a total activation budget, not a stricter prompt.

\paragraph{Anyone who confines an agent by its tool list.}
\S\ref{sec:agent} says that the tool list is a capability map, that the agent's reach is its closure, and that removing tools by name contains an effect only when no remaining tool reaches it.
In the one environment we inventoried, that condition failed for half the effect classes because of a shell.
Whether a shell belongs in the list is a design decision; that the list must be checked by closure rather than by name is not.

\section{What did not work, and what we had to fix to know}
\label{sec:negative}

\begin{itemize}
  \item \textbf{The first figure erased the result.}  We first drew the instantaneous cut as its \emph{maximum} over a period.  For \texttt{rot} the maximum equals the temporal cut, and the two bars were the same height: the gap we were reporting was invisible in our own figure.  The defender does not see the maximum; it sees a random instant.  The figure now shows the median, which for \texttt{rot} is zero.
  \item \textbf{Coincident curves hid the control.}  \texttt{ring} and \texttt{fixed+ring} take the same number of steps, as do \texttt{rot} and \texttt{rot+ring}.  Drawn with the same marker, one of each pair vanished and the legend showed colours that appeared nowhere.  Figure~\ref{fig:N1} uses distinct markers and says in its title which curves coincide.
  \item \textbf{A random shortcut dimension polluted one column.}  The shortcut lattice originally chose its shortcut dimension at random; at one size it chose the top bit, the shortcut itself crossed the split, and the cut column jumped for that size alone.  The dimension is now $\lfloor d/2 \rfloor$, never the top bit.
  \item \textbf{The first day is not a typical day.}  Our first Internet table divided by the first snapshot's cut and reported $\times 2.7$.  The first snapshot happened to be below the median, and the ratio was partly that.  Dividing by the median instantaneous cut gives $\times$\nAsRewireHi{}.
  \item \textbf{Growth is not rewiring.}  A third of the two-year gap in 1997--2000 was new nodes.  Restricting to edges between first-day nodes removed it.
  \item \textbf{The tool's first map was wrong, and the tool said so.}  Every failure of the map against the code was a discovery about the code; the map was corrected until it described what exists, and the discoveries are reported in \S\ref{sec:tool}.
  \item \textbf{The tool itself had two blind spots.}  On a second application it failed to follow any channel sent with \texttt{ipcRenderer.send}, and failed to match any module imported under the \texttt{node:} prefix; it reported three dangerous surfaces where there were \nGTwoReach{}.  The first application used neither feature.  A static analyser is only as complete as the code it was tested on, which is one more instance of the paper's claim.
  \item \textbf{The round-robin schedule reached in about $n/2$ phases, not $\log n$.}  We expected a schedule in which every pair meets once per cycle to reach everything in about $\log_2 n$ phases, as the hypercube does.  In circle-method order it takes about $n/2$ (\nRrSteps{} at $n = \nRrN$), no better than a ring.  The same matchings in random order take \nRrStepsShuf.  Nothing about the cut changes between the two orders.  We had been treating ``how fast it reaches'' and ``how much a snapshot misses'' as one property of a schedule; they are two, and the second is blind to the first.
  \item \textbf{The agent inventory was expected to show a snapshot gap in effects, and showed none.}  We took the inventory of \S\ref{sec:agent} expecting the \nAgDeferred{} deferred tools to hide effects from a start-of-session snapshot, as the rotating dimension hides edges.  They hide \nAgDeferred{} tools and zero effects, because a shell is visible from the start.  The result we had in mind was wrong for that environment; the result that held, that a deny-list leaves \nAgDenyStill{} of \nAgEffects{} effects reachable, is the one \S\ref{sec:tool} had already found for programs.  We report the expectation because it is the kind of reasoning from a snapshot the paper argues against, and we had done it ourselves.
  \item \textbf{Confinement (\S\ref{sec:confine}) is analytic.}  We measured it anyway and it matched to four places.  It contributes nothing new; it is in the paper as a check that the simulator is correct, and we say so.
  \item \textbf{So, in the end, is most of \S\ref{sec:snapshot}.}  We measured the snapshot gap, the window defender and the clock-lag defender before we sat down and derived them (\S\ref{sec:closed}), and the derivations are four lines each.  Had we done the counting first we would have run fewer simulations and understood the result sooner.  We report the order because it is the honest one, and because the agreement between the two is now the strongest evidence that the simulator computes what we claim.
\end{itemize}

\section{Limitations}
\label{sec:limits}

\begin{itemize}
  \item One balanced partition, by the top bit.  Not the minimum cut.  All comparisons between wirings are at equal partition.
  \item Synchronous steps; an adversary that traverses edges and nothing else; three routing orders from one family.
  \item The real-network measurement is of one kind of network, the AS graph, at two epochs.  Enterprise and datacentre networks may rewire faster.  Regions are grown by breadth-first search from random seeds and are not administrative boundaries.
  \item The tool was built for one application and then applied to the desktop shell of a second.  Its source, both maps and its output on both are included with this paper; the applications are private, so the outputs are included as records rather than reproduced.
  \item The round-robin measurement (\S\ref{sec:rotor}) is of the schedule those fabrics share, not of any fabric: one matching per phase, no always-on layer, no traffic.  Opera's expander layer would put a nonzero $|A|$ under it.
  \item The air-gap reading (\S\ref{sec:airgap}) is a derivation and cites the literature for its facts; we measured nothing at an air gap.  The spread-spectrum reading (\S\ref{sec:hopping}) is a translation and claims nothing that field does not already know.
  \item The agent inventory (\S\ref{sec:agent}) is one session of one commercial environment, classified by the agent itself from tool descriptions, and it does not see the permission layer that gates individual calls.  It is a reading of a declared surface, not an assessment of the product, and it is included because it is reproducible in the weak sense that the record is public and can be disputed line by line.
  \item The amnesia inventory (\S\ref{sec:amnesia}) declares the persistence of each effect class from the agent's own reading of its harness; it does not observe an actual reset, and a harness that silently persists working context would change the count.  The spawn model (\S\ref{sec:spawn}) adds nothing to branching theory; its content is the identification of $p$ with a harness's approval rate and of $b$ with what an agent will attempt, and it assumes approvals are independent, which a harness that remembers its refusals would violate.
  \item The capacity reading (\S\ref{sec:capacity}) is an observation about units, not a theorem; we compute no capacities.
  \item All three systems are the author's.  That is the point of the paper and also its weakness: no one else has yet tried to break the argument.
\end{itemize}

\bibliographystyle{plainnat}

\appendix
\section{Reproduction}
\label{app:repro}

Every number in this paper is written by \texttt{figs2.py} into \texttt{tables/} from five record files and one directory of tool output; the prose reads them through macros and contains no typed numbers.

\paragraph{Model.}
\texttt{net/netsim.py} needs only NumPy.  \texttt{--fig} draws Figures~\ref{fig:N1} and~\ref{fig:N2}.
\begin{verbatim}
python net/netsim.py --d 8,10,12,14 --trials 6 --fig
\end{verbatim}
runs in about two minutes on one CPU core and writes \texttt{net/records.json}.

\paragraph{Round-robin schedule.}
\begin{verbatim}
python net/rotor.py --n 256,1024,4096,16384
\end{verbatim}
runs in under a minute and writes \texttt{net/records\_rotor.json}.

\paragraph{Internet.}
The daily record is \texttt{as-733} from the Stanford SNAP collection; the monthly record is CAIDA's \texttt{serial-1} AS-relationships series, files \texttt{20240901} through \texttt{20260901}.
\begin{verbatim}
python net/asreal.py --dir <as-733 directory>
python net/asreal.py --dir <caida directory> --monthly \
                     --out net/records_as_2026.json
\end{verbatim}

\paragraph{Tool.}
\texttt{net/guard/} contains the checker, the two maps and the planted-hole test, copied from the application repository at the commit recorded in \texttt{net/guard/records/PROVENANCE.txt}.
The checker needs the application to run; its output on the application is in \texttt{net/guard/records/} and Tables~\ref{tab:G1} and~\ref{tab:G2} are generated from those files.
In the application repository the commands are \texttt{python -m guard.check} and \texttt{python -m guard.test\_check}.

\paragraph{Agent inventory.}
\texttt{net/agent/tools.json} is the inventory of \S\ref{sec:agent}, with the classification of each tool and the note on which tool is treated as universal; \texttt{net/agent/agentsim.py} computes Table~\ref{tab:A1} from it.
\begin{verbatim}
python net/agent/agentsim.py
\end{verbatim}
Taking a new inventory in another environment means writing another \texttt{tools.json}; the script does not change.
The persistence classification of \S\ref{sec:amnesia} is the \texttt{\_persist} block of the same file; \texttt{net/agent/amnesia.py} computes Table~\ref{tab:M1} from it, and \texttt{net/agent/spawn.py} computes Table~\ref{tab:S1} (NumPy only, about a minute).
\begin{verbatim}
python net/agent/amnesia.py
python net/agent/spawn.py
\end{verbatim}

\paragraph{Paper.}
\begin{verbatim}
python paper2/figs2.py && cd paper2 && latexmk -pdf main.tex
\end{verbatim}

\end{document}

%% file: tables/numbers.tex
\newcommand{\nNBig}{16{,}384}
\newcommand{\nDBig}{14}
\newcommand{\nRotSteps}{14}
\newcommand{\nCubeSteps}{11}

\newcommand{\nSmallSteps}{219.5}
\newcommand{\nFixedRingSteps}{8{,}110}
\newcommand{\nRotZero}{93}
\newcommand{\nRotTemporal}{8{,}192}
\newcommand{\nRlTemporal}{8{,}194}
\newcommand{\nRlInst}{2}
\newcommand{\nRlRatio}{4{,}097}
\newcommand{\nRotAware}{585}
\newcommand{\nAwareRatio}{14}
\newcommand{\nRingTaps}{100}
\newcommand{\nRotTaps}{12}
\newcommand{\nCubeTaps}{12}
\newcommand{\nTapBudget}{2{,}048}
\newcommand{\nRotPq}{68}
\newcommand{\nCubePq}{12}
\newcommand{\nRlPq}{37}
\newcommand{\nRotPqRatio}{4.9}
\newcommand{\nCubePqRatio}{1.1}
\newcommand{\nAsDays}{733}
\newcommand{\nAsFirst}{1997-11-08}
\newcommand{\nAsLast}{2000-01-02}
\newcommand{\nAsNodesA}{3{,}015}
\newcommand{\nAsEdgesA}{5{,}156}
\newcommand{\nAsRewireLo}{1.8}
\newcommand{\nAsRewireHi}{2.0}
\newcommand{\nAsGrowLo}{2.3}
\newcommand{\nAsGrowHi}{2.9}
\newcommand{\nCaMonths}{25}
\newcommand{\nCaFirst}{2024-09-01}
\newcommand{\nCaLast}{2026-09-01}
\newcommand{\nCaNodesA}{77{,}241}
\newcommand{\nCaEdgesA}{493{,}493}
\newcommand{\nCaRewireLo}{1.5}
\newcommand{\nCaRewireHi}{1.6}
\newcommand{\nCaGrowLo}{1.5}
\newcommand{\nCaGrowHi}{1.6}
\newcommand{\nGHoles}{8}
\newcommand{\nGMissed}{6}
\newcommand{\nGCaughtOld}{2}
\newcommand{\nObsOneHeld}{8}
\newcommand{\nObsHalfW}{7}
\newcommand{\nObsHalfHeld}{55}

\newcommand{\nLagZeroHeld}{100}
\newcommand{\nLagOneHeld}{0}
\newcommand{\nPqLowP}{0.05}
\newcommand{\nPqRotLow}{334}
\newcommand{\nPqSmallLow}{3{,}232}
\newcommand{\nPqCubeLow}{30}
\newcommand{\nPqRotSmallRatio}{10}
\newcommand{\nPqRotCubeRatio}{11}
\newcommand{\nGTwoExposed}{38}
\newcommand{\nGTwoReach}{11}
\newcommand{\nGTwoHoles}{2}
\newcommand{\nGTwoCaught}{2}

\newcommand{\nPeriod}{14}
\newcommand{\nCrossR}{8{,}192}
\newcommand{\nAgTools}{143}
\newcommand{\nAgStart}{43}
\newcommand{\nAgDeferred}{100}
\newcommand{\nAgVisiblePct}{30}
\newcommand{\nAgEffects}{10}

\newcommand{\nAgDenyStill}{5}
\newcommand{\nAgDenyLost}{5}
\newcommand{\nAgShellLost}{1}
\newcommand{\nAgShellEffects}{5}
\newcommand{\nAgShellLostName}{read environment secrets}
\newcommand{\nRrN}{16{,}384}
\newcommand{\nRrT}{16{,}383}
\newcommand{\nRrSteps}{8{,}111}
\newcommand{\nRrStepsShuf}{17}
\newcommand{\nRrInstMean}{4{,}096}
\newcommand{\nRrTemporal}{67{,}108{,}864}

\newcommand{\nRrSmallN}{256}
\newcommand{\nRrSmallSteps}{127}
\newcommand{\nRrSmallShuf}{10}
\newcommand{\nRrHeldAlmost}{0}

\newcommand{\nAmIdle}{0}
\newcommand{\nAmChannels}{6}
\newcommand{\nAmTools}{66}
\newcommand{\nAmNetRemoved}{6}
\newcommand{\nAmNetLost}{0}
\newcommand{\nAmFutureTools}{5}
\newcommand{\nSpThreshInv}{8}
\newcommand{\nSpEscAbove}{21}
\newcommand{\nSpEscBelow}{0}
\newcommand{\nSpGensBig}{11}
\newcommand{\nSpCap}{100{,}000}
\newcommand{\nSpBudgetMax}{4{,}096}
\newcommand{\nSpBudgetExtinct}{100}
\newcommand{\nSpInfEscape}{90}

%% file: tables/tabP1.tex
\begin{tabular}{llrr}
\toprule
quantity & closed form & predicted & measured \\
\midrule
instants with no crossing edge (\texttt{rot}) & $(T-1)/T$ & 92.9\% & 92.9\% \\
edges seen at a random instant (\texttt{rot+ring}) & $|A|$ & 2 & 2 \\
edges to block forever (\texttt{rot+ring}) & $|A|+|R|$ & 8194 & 8194 \\
gap ratio (\texttt{rot+ring}) & $(|A|+|R|)/|A|$ & 4097 & 4097 \\
time-aware cost (\texttt{rot}) & $|A|+|R|/T$ & 585.1 & 585.1 \\
time-aware cost (\texttt{rot+ring}) & $|A|+|R|/T$ & 587.1 & 587.1 \\
containment, window $w=1$ & $\min(1,w/T)$ & 7\% & 8\% \\
containment, window $w=2$ & $\min(1,w/T)$ & 14\% & 12\% \\
containment, window $w=4$ & $\min(1,w/T)$ & 29\% & 35\% \\
containment, window $w=7$ & $\min(1,w/T)$ & 50\% & 55\% \\
containment, window $w=14$ & $\min(1,w/T)$ & 100\% & 100\% \\
\bottomrule
\end{tabular}

%% file: tables/tabN1scale.tex
\begin{tabular}{lrrrr}
\toprule
wiring & $n=256$ & $n=1024$ & $n=4096$ & $n=16384$ \\
\midrule
\texttt{ring} & 127 & 507 & 2028 & 8110 \\
\texttt{ring+sc} & 26 & 48 & 108 & 220 \\
\texttt{fixed} & never & never & never & never \\
\texttt{fixed+ring} & 127 & 507 & 2028 & 8110 \\
\texttt{rot} & 8 & 10 & 12 & 14 \\
\texttt{rot+ring} & 8 & 10 & 12 & 14 \\
\texttt{cube} & 7 & 9 & 10 & 11 \\
\bottomrule
\end{tabular}

%% file: tables/tabN1.tex
\begin{tabular}{lrrrrrr}
\toprule
wiring & degree & steps to 99\% & crossing edges (median) & instants at 0 & block forever & traffic seen \\
\midrule
\texttt{ring} & 2 & 8110 & 2 & 0\% & 2 & 100\% \\
\texttt{ring+sc} & 3 & 220 & 2 & 0\% & 2 & 100\% \\
\texttt{fixed} & 1 & never (0\%) & 0 & 100\% & 0 & --- \\
\texttt{fixed+ring} & 3 & 8110 & 2 & 0\% & 2 & 100\% \\
\texttt{rot} & 1 & 14 & 0 & 93\% & 8192 & 12\% \\
\texttt{rot+ring} & 3 & 14 & 2 & 0\% & 8194 & 11\% \\
\texttt{cube} & 14 & 11 & 8192 & 0\% & 8192 & 12\% \\
\bottomrule
\end{tabular}

%% file: tables/tabN5.tex
\begin{tabular}{rrrrrrr}
\toprule
 & \multicolumn{2}{c}{\texttt{rot}} & \multicolumn{2}{c}{\texttt{rot+ring}} & \multicolumn{2}{c}{\texttt{cube}} \\
window $w$ & blocked & contained & blocked & contained & blocked & contained \\
\midrule
1 & 614 & 8\% & 616 & 8\% & 8192 & 100\% \\
2 & 1024 & 12\% & 1026 & 12\% & 8192 & 100\% \\
4 & 2867 & 35\% & 2869 & 35\% & 8192 & 100\% \\
7 & 4506 & 55\% & 4508 & 55\% & 8192 & 100\% \\
14 & 8192 & 100\% & 8194 & 100\% & 8192 & 100\% \\
28 & 8192 & 100\% & 8194 & 100\% & 8192 & 100\% \\
\bottomrule
\end{tabular}

%% file: tables/tabN6.tex
\begin{tabular}{lrrrrrrrrrrrrrrr}
\toprule
clock lag (steps) & 0 & 1 & 2 & 3 & 4 & 5 & 6 & 7 & 8 & 9 & 10 & 11 & 12 & 13 & 14 \\
\midrule
contained & 100\% & 0\% & 0\% & 0\% & 0\% & 0\% & 0\% & 0\% & 0\% & 0\% & 0\% & 0\% & 0\% & 0\% & 100\% \\
\bottomrule
\end{tabular}

%% file: tables/tabN4.tex
\begin{tabular}{lrrrrrrrr}
\toprule
wiring & $p=1$ & $p=0.5$ & $p=0.25$ & asc & desc & random & time-aware (mean) & forever \\
\midrule
\texttt{ring} & 8110 & 16185 & 32657 & 100\% & 100\% & 100\% & 2 & 2 \\
\texttt{ring+sc} & 220 & 389 & 720 & 100\% & 100\% & 100\% & 2 & 2 \\
\texttt{fixed+ring} & 8110 & 13483 & 25605 & 100\% & 100\% & 100\% & 2 & 2 \\
\texttt{rot} & 14 & 32 & 68 & 12\% & 12\% & 12\% & 585 & 8192 \\
\texttt{rot+ring} & 14 & 24 & 37 & 11\% & 10\% & 11\% & 587 & 8194 \\
\texttt{cube} & 11 & 11 & 12 & 12\% & 12\% & 12\% & 8192 & 8192 \\
\bottomrule
\end{tabular}

%% file: tables/tabN2.tex
\begin{tabular}{rrr}
\toprule
$d'$ & measured & $2^{d'-d}$ \\
\midrule
1 & 0.0005 & 0.0005 \\
2 & 0.0010 & 0.0010 \\
3 & 0.0020 & 0.0020 \\
4 & 0.0039 & 0.0039 \\
5 & 0.0078 & 0.0078 \\
6 & 0.0156 & 0.0156 \\
7 & 0.0312 & 0.0312 \\
8 & 0.0625 & 0.0625 \\
9 & 0.1250 & 0.1250 \\
10 & 0.2500 & 0.2500 \\
11 & 0.5000 & 0.5000 \\
\bottomrule
\end{tabular}

%% file: tables/tabN7.tex
\begin{tabular}{lrrrrr}
\toprule
wiring & $p=1$ & $p=0.5$ & $p=0.25$ & $p=0.1$ & $p=0.05$ \\
\midrule
\texttt{ring} & 8110 & 16262 & 32340 & 81347 & 162765 \\
\texttt{ring+sc} & 212 & 346 & 727 & 1796 & 3232 \\
\texttt{rot} & 14 & 32 & 68 & 168 & 334 \\
\texttt{rot+ring} & 14 & 24 & 38 & 86 & 184 \\
\texttt{cube} & 11 & 11 & 12 & 18 & 30 \\
\bottomrule
\end{tabular}

%% file: tables/tabN3.tex
\begin{tabular}{rrrrrrrrrrrr}
\toprule
 & & \multicolumn{5}{c}{rewiring only} & \multicolumn{5}{c}{with growth} \\
region & seen (median) & 30 d & 90 d & 180 d & 365 d & 733 d & 30 d & 90 d & 180 d & 365 d & 733 d \\
\midrule
50 & 1004 & $\times$1.1 & $\times$1.1 & $\times$1.2 & $\times$1.4 & $\times$1.8 & $\times$0.8 & $\times$0.9 & $\times$1.1 & $\times$1.4 & $\times$2.3 \\
200 & 2257 & $\times$1.1 & $\times$1.2 & $\times$1.3 & $\times$1.5 & $\times$1.8 & $\times$0.8 & $\times$0.9 & $\times$1.1 & $\times$1.4 & $\times$2.3 \\
1000 & 2125 & $\times$1.0 & $\times$1.1 & $\times$1.3 & $\times$1.5 & $\times$2.0 & $\times$0.7 & $\times$0.8 & $\times$1.1 & $\times$1.5 & $\times$2.9 \\
\bottomrule
\end{tabular}

%% file: tables/tabN3b.tex
\begin{tabular}{rrrrrrrrrrrr}
\toprule
 & & \multicolumn{5}{c}{rewiring only} & \multicolumn{5}{c}{with growth} \\
region & seen (median) & 3 mo & 6 mo & 12 mo & 18 mo & 24 mo & 3 mo & 6 mo & 12 mo & 18 mo & 24 mo \\
\midrule
50 & 10733 & $\times$1.2 & $\times$1.2 & $\times$1.3 & $\times$1.4 & $\times$1.5 & $\times$1.2 & $\times$1.2 & $\times$1.3 & $\times$1.4 & $\times$1.5 \\
200 & 90166 & $\times$1.1 & $\times$1.2 & $\times$1.3 & $\times$1.4 & $\times$1.5 & $\times$1.1 & $\times$1.2 & $\times$1.3 & $\times$1.5 & $\times$1.6 \\
1000 & 219184 & $\times$1.1 & $\times$1.2 & $\times$1.4 & $\times$1.5 & $\times$1.6 & $\times$1.1 & $\times$1.2 & $\times$1.4 & $\times$1.5 & $\times$1.6 \\
\bottomrule
\end{tabular}

%% file: tables/tabR1.tex
\setlength{\tabcolsep}{3.5pt}
\begin{tabular}{rrrrrrrcc}
\toprule
$n$ & $T$ & steps, circle & steps, shuffled & inst.\ mean & temporal & ratio & held at $w=T-1$ & held at $w=T$ \\
\midrule
256 & 255 & 127 & 10 & 64 & 16{,}384 & 255 & 0\% & 100\% \\
1{,}024 & 1{,}023 & 507 & 13 & 256 & 262{,}144 & 1{,}023 & 0\% & 100\% \\
4{,}096 & 4{,}095 & 2{,}028 & 15 & 1{,}024 & 4{,}194{,}304 & 4{,}095 & 0\% & 100\% \\
16{,}384 & 16{,}383 & 8{,}111 & 17 & 4{,}096 & 67{,}108{,}864 & 16{,}383 & 0\% & 100\% \\
\bottomrule
\end{tabular}

%% file: tables/tabG1.tex
\begin{tabular}{lp{0.52\linewidth}cc}
\toprule
 & hole planted & string deny-list & reachability check \\
\midrule
0 & no hole planted & passes & passes \\
1 & call the order function directly from the chat entry & \textbf{missed} & caught \\
2 & call it through an alias not in the deny list & \textbf{missed} & caught \\
3 & reach it through a helper the entry already calls (one hop) & \textbf{missed} & caught \\
4 & call \texttt{.place()} on an object of unknown type & caught & caught \\
5 & write the paper-account ledger file directly (no buy function) & \textbf{missed} & caught \\
6 & expose a seventh function from the desktop preload & caught & caught \\
7 & open an external URL from the main-process handler of save & \textbf{missed} & caught \\
8 & spawn a command inside the settings module that save calls & \textbf{missed} & caught \\
\bottomrule
\end{tabular}

%% file: tables/tabG2.tex
\begin{tabular}{llrrrr}
\toprule
application & side & modules/files & functions & edges & unresolved calls \\
\midrule
trading & JavaScript (desktop) & 5 & 46 & 124 & 71 \\
trading & Python (application) & 80 & 534 & 1328 & 3243 \\
note-taking & JavaScript (desktop) & 2 & 99 & 256 & 205 \\
\bottomrule
\end{tabular}

%% file: tables/tabA1.tex
\begin{tabular}{lrrccl}
\toprule
effect & at start & deferred & reachable at start & after deny-list & via \\
\midrule
write a file in the sandbox & 3 & 3 & yes & yes & Bash \\
start a process & 1 & 3 & yes & yes & Bash \\
reach the network & 1 & 6 & yes & yes & Bash \\
change the repository & 1 & 7 & yes & yes & Bash \\
change an external service & 12 & 28 & yes & no & -- \\
publish to a public place & 1 & 9 & yes & no & -- \\
reach a person & 7 & 4 & yes & no & -- \\
start another agent & 6 & 0 & yes & no & -- \\
act at a future time & 4 & 1 & yes & no & -- \\
read environment secrets & 1 & 0 & yes & yes & Bash \\
\bottomrule
\end{tabular}

%% file: tables/tabM1.tex
\begin{tabular}{llr}
\toprule
persistence & effect classes & tools \\
\midrule
leaves no state & exec, net, read, secret\_read & 72 \\
survives within the session & file\_write & 5 \\
delivered outside the boundary & message & 10 \\
survives any reset (commit, external, publish) & ext\_write, publish, repo\_write & 46 \\
reconstructs a future activation & schedule & 4 \\
creates another agent & spawn & 6 \\
\bottomrule
\end{tabular}

%% file: tables/tabS1.tex
\begin{tabular}{rrrrr}
\toprule
$b$ & $p$ & $m=bp$ & escape (measured) & escape (branching) \\
\midrule
2 & 0.250 & 0.50 & 0.000 & 0.000 \\
2 & 0.450 & 0.90 & 0.000 & 0.000 \\
2 & 0.500 & 1.00\;$\leftarrow$ & 0.000 & 0.019 \\
2 & 0.550 & 1.10 & 0.329 & 0.331 \\
2 & 0.750 & 1.50 & 0.888 & 0.889 \\
4 & 0.125 & 0.50 & 0.000 & 0.000 \\
4 & 0.225 & 0.90 & 0.000 & 0.000 \\
4 & 0.250 & 1.00\;$\leftarrow$ & 0.000 & 0.013 \\
4 & 0.275 & 1.10 & 0.228 & 0.230 \\
4 & 0.375 & 1.50 & 0.720 & 0.710 \\
4 & 0.750 & 3.00 & 0.997 & 0.996 \\
8 & 0.062 & 0.50 & 0.000 & 0.000 \\
8 & 0.113 & 0.90 & 0.000 & 0.000 \\
8 & 0.125 & 1.00\;$\leftarrow$ & 0.000 & 0.011 \\
8 & 0.138 & 1.10 & 0.209 & 0.199 \\
8 & 0.188 & 1.50 & 0.651 & 0.641 \\
8 & 0.375 & 3.00 & 0.974 & 0.974 \\
\bottomrule
\end{tabular}